\documentclass[twocolumn]{aastex701}

\usepackage{multirow}
\usepackage{enumerate}
\usepackage{amsmath}
\usepackage{amssymb} 
\usepackage[nointegrals]{wasysym}
\shorttitle{Exoring Reflected Light Phase Curves}
\shortauthors{Lane et al.}
\graphicspath{{./}{figures/}}

\begin{document}
\title{Detecting Exoplanetary Rings with Roman CGI 
Reflected Light Phase Curve Observations}
\correspondingauthor{Elizabeth Lane}
\email{laneeli@umich.edu}

\author[0009-0007-5843-6447]{Elizabeth Lane}
\affiliation{Department of Astronomy, University of Michigan, Ann Arbor, MI 48109, USA}
\email{laneeli@umich.edu}

\author[0000-0002-9521-9798]{Mary Anne Limbach}
\affiliation{Department of Astronomy, University of Michigan, Ann Arbor, MI 48109, USA}
\email{mlimbach@umich.edu}

\author[0000-0003-3904-7378]{Logan Pearce}
\affiliation{Department of Astronomy, University of Michigan, Ann Arbor, MI 48109, USA}
\email{elapearce@umich.edu}

\author[0000-0002-4894-193X]{Samantha Hasler}
\affiliation{{Space Telescope Science Institute, 3700 San Martin Drive, Baltimore, MD 21218, USA}}
\email{shasler@stsci.edu}

\author[0000-0002-7162-8036]{Alexandra Z. Greenbaum}
\affiliation{IPAC, Mail Code 100-22, Caltech, 1200 E. California Boulevard, Pasadena, CA 91125, USA}
\email{azg@ipac.caltech.edu}


\begin{abstract}

    The \textit{Nancy Grace Roman Space Telescope}'s Coronagraph Instrument (CGI) will provide the first opportunity to directly detect mature giant exoplanets in reflected light. However, most known radial-velocity planets lie near CGI's sensitivity limit. In contrast, Saturn-like rings around these planets may be much more accessible: our models show that rings can outshine their host planets by one to two orders of magnitude or more, making them easier to detect than the planets themselves. However, because the planet and rings would be unresolved in reflected light, unrecognized rings could lead to misinterpretation of the planet’s intrinsic brightness and physical properties. To disentangle these blended signals, we model reflected-light phase curves of ringed and ringless planets. Ring systems can produce a distinctive phase-curve morphology, often with four extrema rather than the two expected for unringed planets, enabling robust confirmation with multi-epoch CGI observations. We develop an analytical framework for ringed-planet phase curves using a modern, self-consistent coordinate system and use it to illustrate phase curves for a range of ring configurations. We then simulate observations of $\epsilon$~Eridani~b (3.5 AU, 1 M\textsubscript{Jup}), a prime initial CGI target, with a Saturn-like ring system. These simulations show that such a configuration would be detectable and suggest that rings may expand the pool of viable CGI targets to planets previously considered beyond reach. We conclude that CGI is poised to provide the first observational constraints on the prevalence of Saturn-like rings around giant exoplanets.

\end{abstract}


\section{Introduction} \label{sec:intro}

    Although great strides have been made to characterize exoplanets’ dynamical and physical properties \citep{2015ARA&A..53..409W,2021ARA&A..59..291Z,2024RvMG...90..411K,2024arXiv240504587F}, there have yet to be any confirmed detections of {mature }exoplanetary rings \citep[exorings;][]{teachey2024detectingcharacterizingexomoonsexorings}. In the Solar System, all four giant planets possess ring systems \citep{1977Natur.267..328E, 1986Natur.319..636H, doi:10.1126/science.246.4936.1422}, suggesting rings may be a common occurrence around mature planets. Therefore, the absence of confirmed exorings likely reflects the observational challenges in detecting and confirming such systems, rather than indicating they are uncommon outside our Solar System. 

    The discovery of exorings would provide important insight into the formation and evolution of planetary systems. Exorings may form through a variety of processes, including collisions between minor bodies \citep{2015NatGe...8..686H}, the tidal disruption of a moon \citep{article, doi:10.1126/science.abn1234}, the partial dissipation of a circumplanetary disk \citep{2003Icar..163..198M,BOTTKE2005111}, or the ejection of material from a moon \citep{doi:10.1126/science.204.4396.951,2018prs..book..517C}. Detecting such systems would therefore offer a window into the dynamical histories of exoplanetary systems. Moreover, characterizing ring properties (e.g., icy versus rocky albedos) could provide insight into ring composition and origin \citep{2007MNRAS.377.1588H}, and potentially trace back to the conditions and locations of planet and satellite formation \citep{2025SSRv..221...66C}.

    Although no ring systems around mature planets have been confirmed, multiple circumplanetary disks around young, forming planets have been discovered \citep{Kenworthy_2015, Benisty_2021, Ohno_2022, Hoch_2025}. In addition, some ``super-puff'' planets with unusually large radii relative to their masses may be explained by the presence of rings \citep{2017AJ....153..193A,Liang_2021, lu2024dynamicalhistoryhip41378f, Akinsanmi_2020, Piro_2020}. In such cases, radii inferred from transit observations may be overestimated, as small planets with extended ring systems can be mistaken for larger, ringless planets \citep{Arnold_2004, Barnes_2004}.

    Many current exoring detection methods rely on transmission spectroscopy and polarimetry, which can reveal how rings flatten spectral features or introduce polarization signatures \citep{2022ApJ...930...50O, Lietzow_2023, Veenstra_2025}. It may also be possible to confirm exorings around ``super-puffs'' with very high-precision follow-up transit observations as a ring would alter the observed transit signature \citep{2004ApJ...616.1193B,2015ApJ...803L..14Z,ZULUAGA2022100623}, for which upcoming JWST observations will test for HIP-41378\,f \citep{2026jwst.prop.11253}. Additional scattered light could also reveal the presence of a non-transiting ring \citep{Sucerquia_2020}. 

    In reflected light, it may be possible to directly image starlight reflected by exoring systems. Saturn’s rings, for example, outshine the planet at most viewing geometries and wavelengths, though they are observed nearly edge-on from Earth. In CH$_4$ absorption bands, Saturn’s rings are more than 100 times brighter than the planet when viewed face-on \citep{miller2023compositionsaturnsrings, doi:10.1126/science.204.4396.951}. Only ring systems with emitting areas much larger than Saturn's rings could be detected with modern observatories \citep[e.g., with JWST NIRCam coronagraphy;][]{bowensrubin2025detectionexoringsreflectedlight}. Furthermore, single-epoch imaging suffers from degeneracies between planet radius, orbital separation, and albedo in the presence of rings, making these properties difficult to disentangle. Even when the orbit of a known planet is well constrained, albedo and other physical properties remain challenging to separate from the contribution of a ring in direct imaging observations \citep{Arnold_2004,2017PASP..129c4401N}. 

    One approach, developed by \cite{Arnold_2004}, takes advantage of the variability of an exoplanet’s reflected flux with orbital phase. Their model predicts how a ring might affect the planet’s reflected light phase curve as a function of ring and planet orientation and viewing angle. At the time of that work, no planned or existing observatories had the sensitivity required to detect such signals. However, with the launch of the \textit{Nancy Grace Roman Space Telescope} \citep{2015arXiv150303757S,2019arXiv190205569A} and its Coronagraph Instrument \citep[CGI;][]{2020SPIE11443E..1UK,2022SPIE12180E..1WM,2024ESS.....520203W}, this is beginning to change. CGI will, for the first time, have the sensitivity to detect {mature} gas giant exoplanets in visible light, with a flux ratio detection limit of $10^{-8}$ or lower \citep{2023SPIE12680E..0TB, zellem2022nancygraceromanspace}, potentially enabling the direct imaging of exoplanets, and their rings, in reflected light.
    
    Here, we present models for reflected light CGI phase curves for ringed and ringless planets. In Section \ref{sec:methods}, we describe the sky and orbit coordinate systems and reference frames, and define the physical parameters that govern the phase curve models. In Section \ref{sec:results}, we present phase curve models for ringed and ringless planets, exploring the impact of ring orientation on the reflected-light phase curve. In Section \ref{sec:epsErib}, we use these models to create simulated data of potential CGI target $\epsilon$~Eridani~b \citep{Carri_n_Gonz_lez_2021, 2024SPIE13092E..55W} with and without a Saturn-like ring system. In Section \ref{sec:discuss}, we compare our model with previous work, discuss the implications for detectability with Roman CGI, and describe our plans for future work. Finally, we conclude with the expected impact and implications of ringed planets on CGI observations in Section \ref{sec:conclusion}.

\section{Methods} \label{sec:methods}

    The reflected light phase curves of ringed and ringless planets can be described with a combination of three analytical functions. These functions depend on planet position, ring orientation, and the planet and ring's optical properties and size. In this section, we discuss the ring and planet orientation parameters, reference frames, and the ringed and ringless planet models. 

\subsection{Orientation Parameters}\label{subsec:orientation_params}

\begin{figure}
\centering
\includegraphics[width=0.45\textwidth]{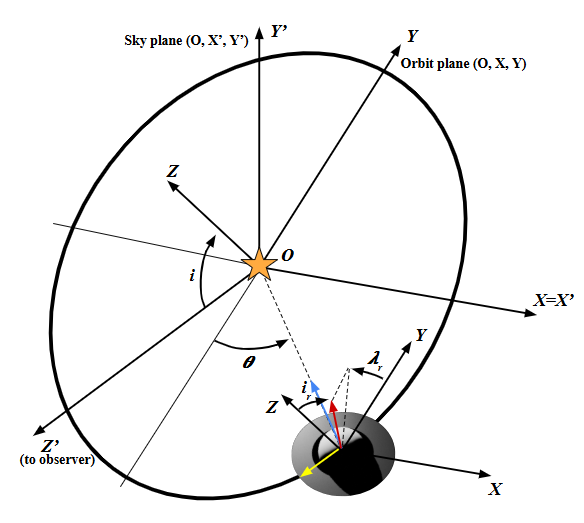}
    \caption{Definition of reference frames and angular parameters. Orbital phase and ring inclination longitude are positive in the counterclockwise direction. Obliquity is defined as the angle between the ring normal (red arrow) and the orbit plane normal (the Z-axis). The inclination is the angle between the sky plane normal and the orbit plane normal. The pictured orbit inclination is negative. Originating from the planet, the red arrow is the ring normal $\vec{n}$, the blue arrow is the vector {$\vec{r}$} pointing toward the star, and the yellow arrow is the vector $\vec{o}$ pointing toward the observer. In the above orientation, the planet's ring is partially shadowed by the planet. This effect is ignored in our models as our goal is to create a first-order approximation.}
\label{ReferenceFrames}
\end{figure}

    The reflected light contrast of a ringed planet is determined by four orientation angles defined in the sky and orbit planes, illustrated in Figure \ref{ReferenceFrames} and defined in Table \ref{tab:orientation_params}. Here, the sky plane is denoted with (X$^\prime$, Y$^\prime$, Z$^\prime$) coordinates, and the orbit plane is denoted with (X, Y, Z) coordinates. The orbit plane is described with the classic Campbell elements: the eccentricity $e$, the semi-major axis $a$, the longitude of ascending node $\Omega$, and the inclination $i$. We assume a circular orbit ($e=0$) and $\Omega=0^\circ$ for simplicity. Under these assumptions, the planet-star separation is constant and can be used in place of the semi-major axis.

    The first orientation angle describing the ringed planet contrast, the inclination $i$, is a rotation of the orbit plane around the line of nodes (the X=X$^\prime$ axis). The angle is measured between the orbit plane normal (the Z-direction) and the sky plane normal (the Z$^\prime$-direction). 

    Second, the orbital phase $\phi$ describes the planet's angular position in its orbit following the modern definition from \cite{2017A&A...599A..16L}. Unlike the definition in \cite{Arnold_2004}, 0$^\circ$ orbital phase is defined when the planet is at inferior conjunction. Therefore, 0$^\circ$ orbital phase occurs when the planet is along the -Y-direction and reflecting the most light. Following this definition, 0$^\circ$ is full phase, 0--90$^\circ$ is gibbous, 90$^\circ$ is quadrature, 90--180$^\circ$ is crescent, and 180$^\circ$ is new phase.

    The orientation of the ring is defined by two angles with respect to the orbit plane. The obliquity $i_r$ is the angle between the ring normal (red arrow in Figure \ref{ReferenceFrames}) and the Z-direction.  When the obliquity is 0$^\circ$, the ring lies in the orbit plane. The ring inclination longitude $\lambda_r$ describes a rotation of the ring about the orbit normal (the Z-direction). Unlike the \cite{Arnold_2004} definition, this angle is measured between the projection of the ring normal onto the orbit plane and the Y-direction, so at $0^\circ$ ring inclination longitude, the planet and ring contrast maxima align.

\begin{table}
\centering
    \caption{Geometric and optical parameters of the ringed planet model following the classic Campbell elements and \cite{Arnold_2004}.} \label{tab:orientation_params}
\medskip
\begin{tabular}{cl}
\hline
Parameter & Definition \\
\hline
\multicolumn{2}{l}{\textbf{Orbit parameters}} \\
        $\phi$ & Orbital phase \\
        $\omega_p$ & Argument of periastron \\
        $i$ & Orbit inclination \\
        {$a$} & {Semi-major axis$^{a}$} \\
        {$r$} & Planet-star separation$^{a}$ \\
        $e$ & Eccentricity$^{a}$ \\
        $\Omega$ & Longitude of the ascending node$^{a}$ \\
\multicolumn{2}{l}{\textbf{Planet parameters}} \\
        $r_p$ & Planet radius \\
        $A_p$ & Planet geometric albedo \\
\multicolumn{2}{l}{\textbf{Ring parameters}} \\
        $r_i$ & Ring inner radius \\
        $r_o$ & Ring outer radius \\
        $\omega_0$ & Ring single scattering albedo \\
        $\tau$ & Normal optical thickness \\
        $i_r$ & Obliquity \\
        $\lambda_r$ & Ring inclination longitude \\
\hline
\end{tabular}
        \tablenotetext{a}{In this work, we assume $e=0$ and $\Omega=0^\circ$. {Because we assume a circular orbit, the semi-major axis $a$ and planet-star separation $r$ are equivalent.}}
\end{table}

\subsection{Ringless Planet Model}\label{reduce}

    Our modeled reflected light phase curves of a ringless planet depend on the planet's geometric albedo, ratio of radius to planet-star separation, and a Lambert phase function, which describes how the planet’s illumination varies with orbital phase. In addition, we assume the planet reflects isotropically, consistent with a Lambertian surface. The full function follows that described by \cite{Cahoy_2010},

\begin{equation}
C_{planet}=A_p \left( \frac{r_p}{r} \right) ^2 \left( \frac{\sin\alpha+(\pi-\alpha)\cos\alpha}{\pi} \right),
\label{eq: Cahoy}
\end{equation}

    \noindent where $A_p$ is the planet's geometric albedo, $r_p$ is the planet radius, {$r$} is the planet-star separation, and $\alpha$ is the phase angle. Similar to \cite{Lacy_2019}, we define the phase angle as

\begin{equation}
\alpha=\cos^{-1}(\sin i\cdot\sin(\phi+\omega_p)),
\label{eq: PhaseAngle}
\end{equation}

    \noindent where $i$ is the orbit inclination, $\phi$ is the orbital phase, and $\omega_p$ is the argument of periastron.

\subsection{Ringed Planet Model}\label{subsec:ringedplanet}

    Here we build upon the work of \cite{Arnold_2004} to present a simple phase curve model as a first-order test of Roman’s ability to characterize exorings. Therefore, for the ringed planet model, we assume the ring is illuminated by parallel stellar radiation (i.e., the star is treated as a distant point source) and scatters as a gray, non-Lambertian surface. We adopt a single-scattering approximation and assume that the ring is {geometrically }thin and homogeneous. We also neglect mutual illumination and shadowing between the ring and planet, which are expected to modify the total reflected ring flux {of a Saturn-like ring ($\tau=1$)} at only the $\sim$5--20\% level. {If the optical depth of the ring increased ($\tau>\approx3$), the shadowing effect would significantly decrease the contrast for half of the orbital period. While mutual illumination may increase the contrast if both the ring and planet have high albedos, the effect is not large enough to counteract the shadowing effect \citep{Arnold_2004}.}

    We use the orientation parameters listed in Table \ref{tab:orientation_params} and described in Section \ref{subsec:orientation_params} to define three vectors that describe the position of the ring in the orbit plane with respect to the star and the observer. The vector normal to the ring (red arrow in Figure \ref{ReferenceFrames}) is given by

\begin{equation}
\vec{n}=
\begin{pmatrix} 
-\sin{i_r}\sin{\lambda_r} \\ 
\sin{i_r}\cos{\lambda_r} \\ 
\cos{i_r}
\end{pmatrix},
\label{eq: Rvec}
\end{equation}

    \noindent where $i_r$ is the obliquity and $\lambda_r$ is the ring inclination longitude. The vector pointing from the ring to the star (blue arrow in Figure \ref{ReferenceFrames}) is

\begin{equation}
\vec{r}=
\begin{pmatrix} 
-\sin{\phi} \\ 
\cos{\phi} \\ 
0
\end{pmatrix},
\label{eq: Nvec}
\end{equation}

    \noindent where $\phi$ is the orbital phase. The vector pointing from the ring to the observer (yellow arrow in Figure \ref{ReferenceFrames}) is

\begin{equation}
\vec{o}=
\begin{pmatrix} 
0 \\ 
\sin{i} \\ 
\cos{i}
\end{pmatrix},
\label{eq: Ovec}
\end{equation}

    \noindent where $i$ is the inclination.

    With these vectors, we define the ring illumination angle $\alpha_1$ as the angle between $\vec{n}$ and {$\vec{r}$}:

\begin{equation}
\alpha_1=\cos^{-1}(\vec{n} \cdot \vec{r}).
\label{eq: alpha_1}
\end{equation}

\begin{figure}
\centering
\includegraphics[width=0.47\textwidth]{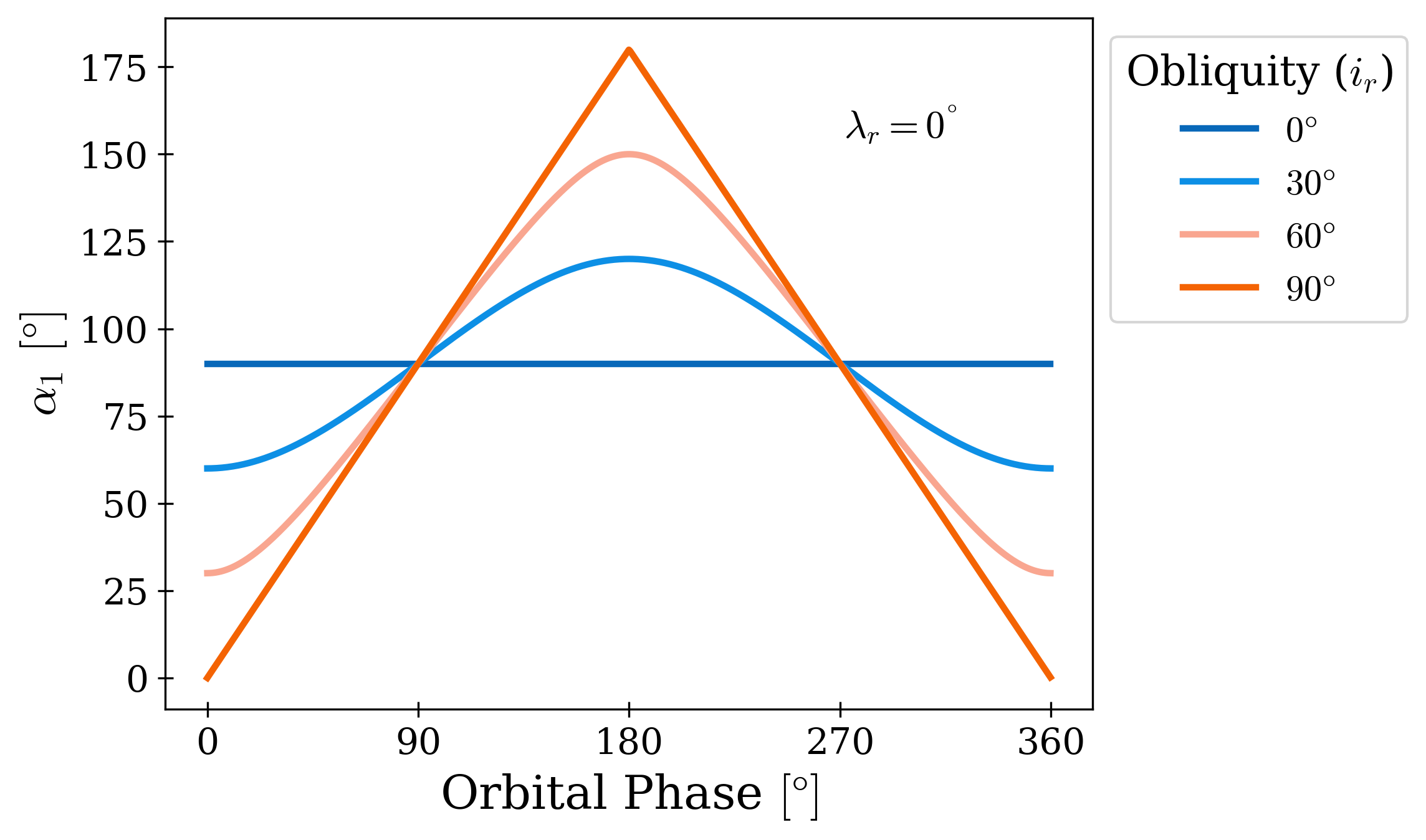}
    \caption{The illumination angle, the angle between the vector normal to the ring's surface and the vector pointing from the ring to the star, as a function of orbital phase. When the ring lies in the orbital plane ($i_r=0^\circ$), these vectors are perpendicular at all points in the orbit. At greater obliquities, the range of illumination angles increases.}
\label{fig:alpha_1}
\end{figure}

    \noindent The variation of the illumination angle with orbital phase $\phi$ is shown in Figure~\ref{fig:alpha_1} for $\lambda_r = 0^\circ$. As shown, the illumination angle $\alpha_1$ varies with $\phi$, with the amplitude of this variation set by the obliquity $i_r$. For $i_r = 0^\circ$, $\alpha_1$ remains fixed at $90^\circ$ for all orbital phases, since the ring normal is perpendicular to the star-ring direction. In contrast, for $i_r = 90^\circ$, $\alpha_1$ varies over the full range from $0^\circ$ to $180^\circ$ throughout the orbit. At intermediate obliquities, $\alpha_1$ varies sinusoidally (except at extreme obliquities) with $\phi$, with increasing amplitude as $i_r$ increases. The extrema in $\alpha_1$ occur when the projection of the ring normal aligns with the star-ring direction.

    Similarly, we define the observer angle $\alpha_2$ as the angle between $\vec{n}$ and $\vec{o}$:

\begin{equation}
\alpha_2=\cos^{-1}(\vec{n} \cdot \vec{o}).
\label{eq: alpha_2}
\end{equation}

    \noindent Throughout the remainder of this work, we will use the values $\mu_1=\cos{\alpha_1}$ and $\mu_2=\cos{\alpha_2}$ to describe the ring's phase curve.

    The ring's phase curve is composed of reflected light when $\mu_1\mu_2 > 0$. While $\mu_1\mu_2 \leq 0$, the ring is reflecting light away from the observer, and no reflected flux is measured. \cite{Arnold_2004} determine that while $\mu_1\mu_2 > 0$, the fraction of incident flux reflected is

\begin{equation}
I_R =
\frac{\omega_0 |\mu_1|}{|\mu_1|+|\mu_2|} \left[ 1-\exp \left( -\tau_{path} \right) \right],
\label{eq: ReflectedLight}
\end{equation}

    \noindent where $\tau_{path}=\tau \left( \frac{1}{|\mu_1|}+\frac{1}{|\mu_2|} \right)$.

    If the ring is not opaque, the phase curve will be impacted by transmitted light when $\mu_1\mu_2 \leq 0$. Therefore, while the ring is reflecting light ( $\mu_1\mu_2 > 0$), transmitted light will not affect the phase curve. If light is transmitted, the fraction of incident flux transmitted is given by

\begin{equation}
I_T = \begin{cases}
\frac{\omega_0 |\mu_1|}{|\mu_2|-|\mu_1|} \left[ \exp \left( \frac{-\tau}{|\mu_2|} \right) - \exp \left(\frac{-\tau}{|\mu_1|} \right) \right] & \text{if } |\mu_1| \neq |\mu_2| \\
\frac{\omega_0 \tau}{|\mu_1|} \exp \left( \frac{-\tau}{|\mu_1|} \right) & \text{if } |\mu_1| = |\mu_2| .
\end{cases}
\label{eq: TransmittedLight}
\end{equation}

    The ring-to-star contrast is then obtained by scaling the appropriate phase-dependent ring intensity by the projected ring area:

\begin{equation}
C_{ring}=\frac{r_o^2-r_i^2}{4r^2}(I_R+I_T),
\label{eq: RingTotal}
\end{equation}

    {\noindent where $r_o$ and $r_i$ are the ring's outer and inner radii.}

    Because $I_R$ and $I_T$ are defined piecewise by the viewing geometry, the observed ring signal is reflected- or transmitted-light dominated at a given orbital phase. The blended ringed-planet phase curve is then the sum of the planet-to-star contrast from Equation~\ref{eq: Cahoy} and the ring-to-star contrast from Equation~\ref{eq: RingTotal}.

\section{Ringed Planet Phase Curves} \label{sec:results}

    After constructing the ringed phase curve model, we explore how the ring and orbital orientations impact the resulting phase curves. 

    The most notable impact of a ring on the phase curve models is the distinctive structure, which exhibits four extrema, compared to the two in the ringless phase curve. One maximum occurs when the ring reflects stellar light, while the second occurs when it transmits light. The two minima arise when the ring is edge-on to the star, contributing little reflected or transmitted light. The presence of a ring also increases the apparent brightness compared to a ringless planet. 
    
\begin{figure}
\centering
\includegraphics[width=0.45\textwidth]{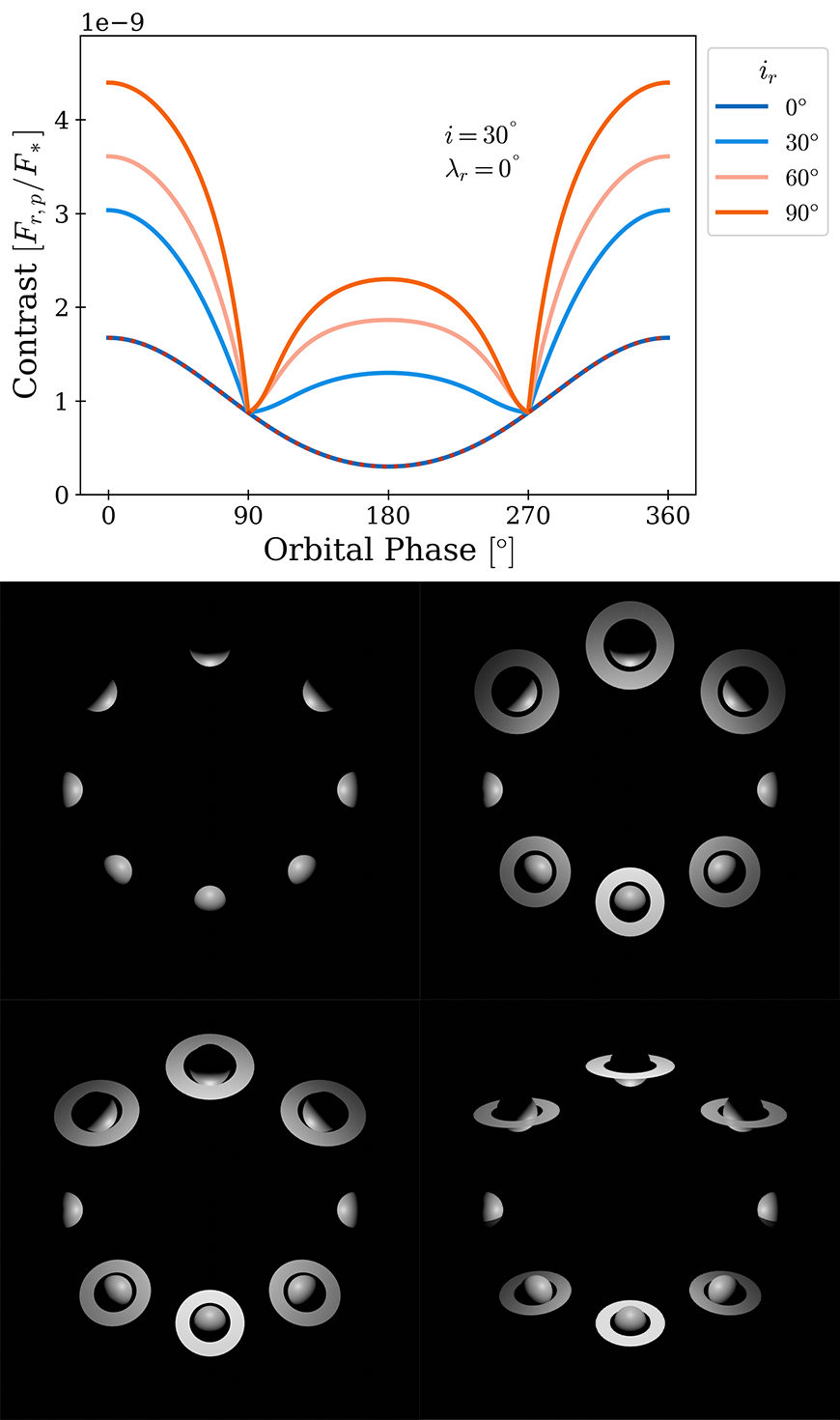}
    \caption{Top: Reflected light phase curves computed with approximate Saturn parameters (Table \ref{parameters}) for a range of obliquities relative to the orbit plane for constant values of orbit inclination and ring inclination longitude. The dotted red line represents the ringless planet, and the solid lines represent the ringed planet phase curves. Bottom: Ring configurations plotted above, rendered as they would appear to an observer using the Blender 3D creation software \citep{Blender}. Displayed obliquities are $i_r=0^\circ$ (top left), $i_r=30^\circ$ (top right), $i_r=60^\circ$ (bottom left), and $i_r=90^\circ$ (bottom right). The orbital phase is indicated by the planet's position, with $0^\circ$ at the bottom and increasing counterclockwise around a central star (not pictured). The ring's obliquity causes the distinctive phase curve with four extrema, except when the obliquity is $0^\circ$. At higher obliquities, the maximum contrast increases and the ring's signature becomes more pronounced. Brightnesses in the illustrations are not precise and should only be used as a visualization of the corresponding ring parameters. Some orientations shown in the illustrations display mutual shadowing{ and illumination}, but this effect is not included in the {model or the }phase curves above.}
\label{Obliquity}
\end{figure}

\begin{figure}
\centering
\includegraphics[width=0.454\textwidth]{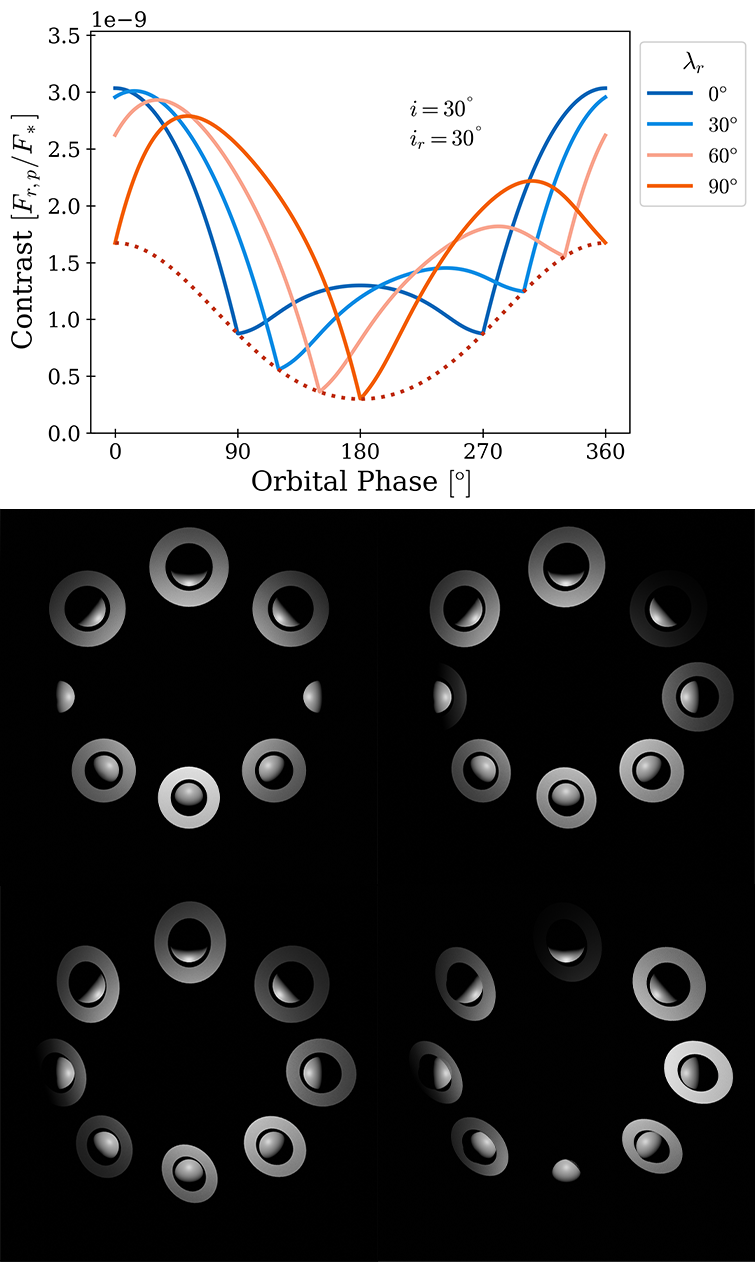}
    \caption{Same as Figure \ref{Obliquity}, showing changing ring inclination longitude for constant orbit inclination and ring obliquity. {The dotted red line represents the ringless planet, and the solid lines represent the ringed planet phase curves. }Displayed ring inclination longitudes are $\lambda_r=0^\circ$ (top left), $\lambda_r=30^\circ$ (top right), $\lambda_r=60^\circ$ (bottom left), and $\lambda_r=90^\circ$ (bottom right). By altering the ring inclination longitude, the ring's contribution to the contrast is shifted along the planet's phase curve. Consequently, the planet and ring extrema may not happen concurrently. {Some orientations shown in the illustrations display mutual shadowing and illumination, but this effect is not included in the model or the phase curves above.}
}
\label{IncLongitude}
\end{figure}

\begin{figure*}
\centering
\includegraphics[width=.85\textwidth]{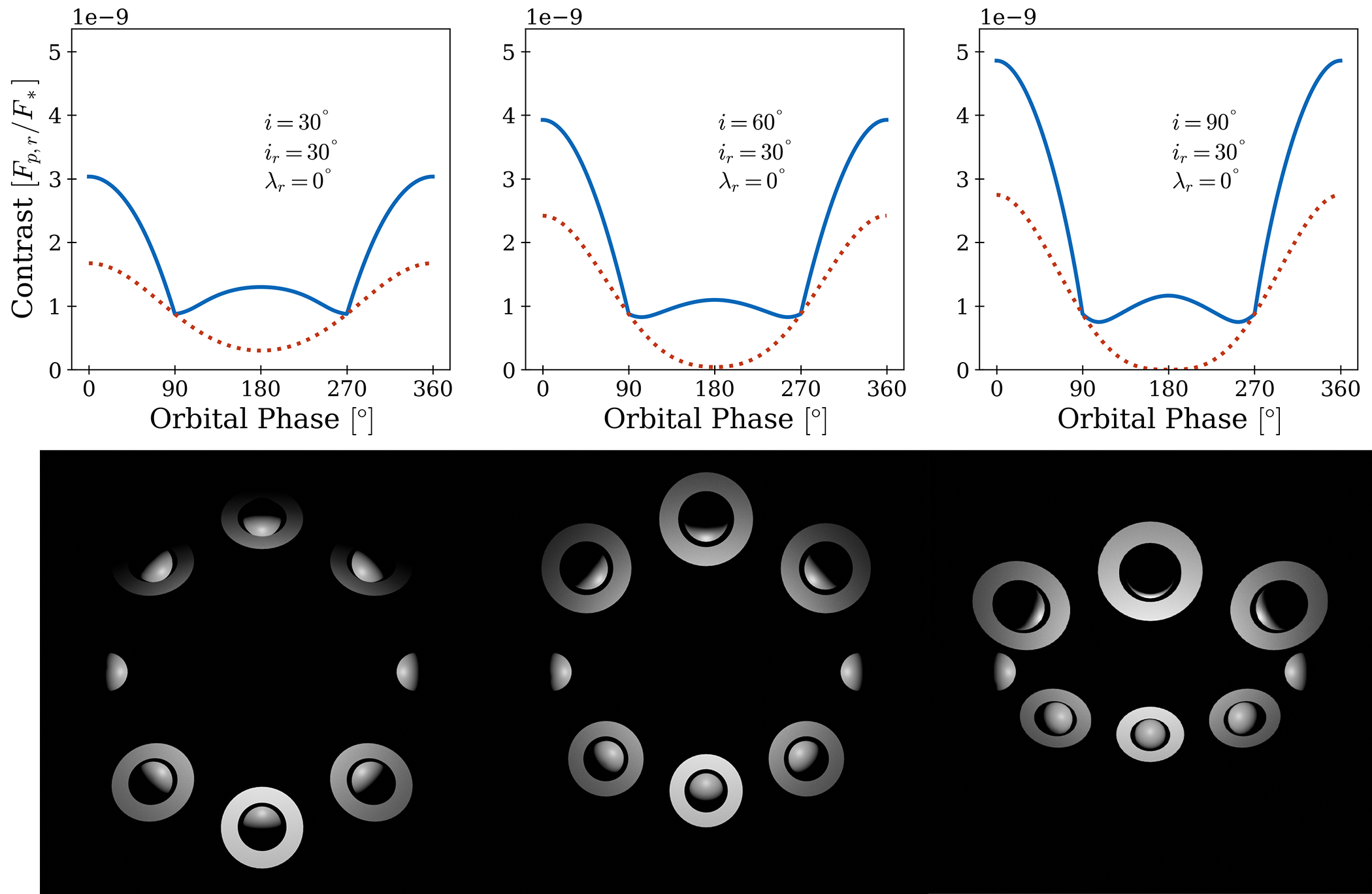}
    \caption{Same as Figure \ref{Obliquity}, showing changing orbital inclinations. {The dotted red line represents the ringless planet, and the solid lines represent the ringed planet phase curves. }Displayed inclinations are $i=30^\circ$ (left), $i=60^\circ$ (middle), and $i=90^\circ$ (right). Increasing the inclination results in an amplification of the planet and ring contrasts. For a face-on orbit ($i=0^\circ$), the planet is always observed at quadrature, producing a flat phase curve. As long as the ring is oblique to the orbital plane ($i_r \neq 0^\circ$), the ring signature will appear regardless of orbital inclination. {Some orientations shown in the illustrations display mutual shadowing and illumination, but this effect is not included in the model or the phase curves above.}}
\label{Inclination}
\end{figure*}

    Figures \ref{Obliquity}, \ref{IncLongitude}, and \ref{Inclination} show the effects of the obliquity, ring inclination longitude, and orbit inclination, respectively. For each phase curve, we show corresponding Blender \citep{Blender} illustrations of the ring system, as viewed by the observer, throughout the orbital phase; these renderings are intended as qualitative visual guides and are not photometrically or geometrically exact.
    
    The strongest impact on the phase curve arises from the ring's obliquity. As shown in Figure \ref{Obliquity}, increasing obliquity leads to the largest peak-to-trough amplitude change compared to the effects of the inclination and ring inclination longitude. When the ring lies in the orbital plane ($i_r=0^\circ$), it is always edge-on to the star and therefore contributes negligible reflected or transmitted light, making this configuration indistinguishable from the ringless case. However, if the obliquity is non-zero, the contrast will vary over the full phase curve and the characteristic four extrema structure emerges. One maximum occurs when the illuminated face of the ring is visible, maximizing reflected light. A second maximum arises when the observer views the ring from the opposite side of the star, allowing transmitted or forward-scattered light to contribute significantly to the observed flux.

    The ring inclination longitude shifts the phase curve extrema along the orbit. As shown in Figure \ref{IncLongitude}, when the ring inclination longitude is aligned with the Y-direction ($\lambda_r=0^\circ$), the maxima align with the extrema in the ringless case. When the ring inclination longitude is non-zero, the ring contrast extrema no longer align with the planet contrast extrema.

    The orbit inclination has a similar effect on ringed planets and ringless planets. Shown in Figure \ref{Inclination}, increasing the inclination increases the maximum contrast. However, unlike a ringless planet, a ringed planet can exhibit significant contrast variation throughout its orbit, even if the orbit is face-on ($i=0^\circ$). 

\begin{figure}
\centering
\includegraphics[width=0.454\textwidth]{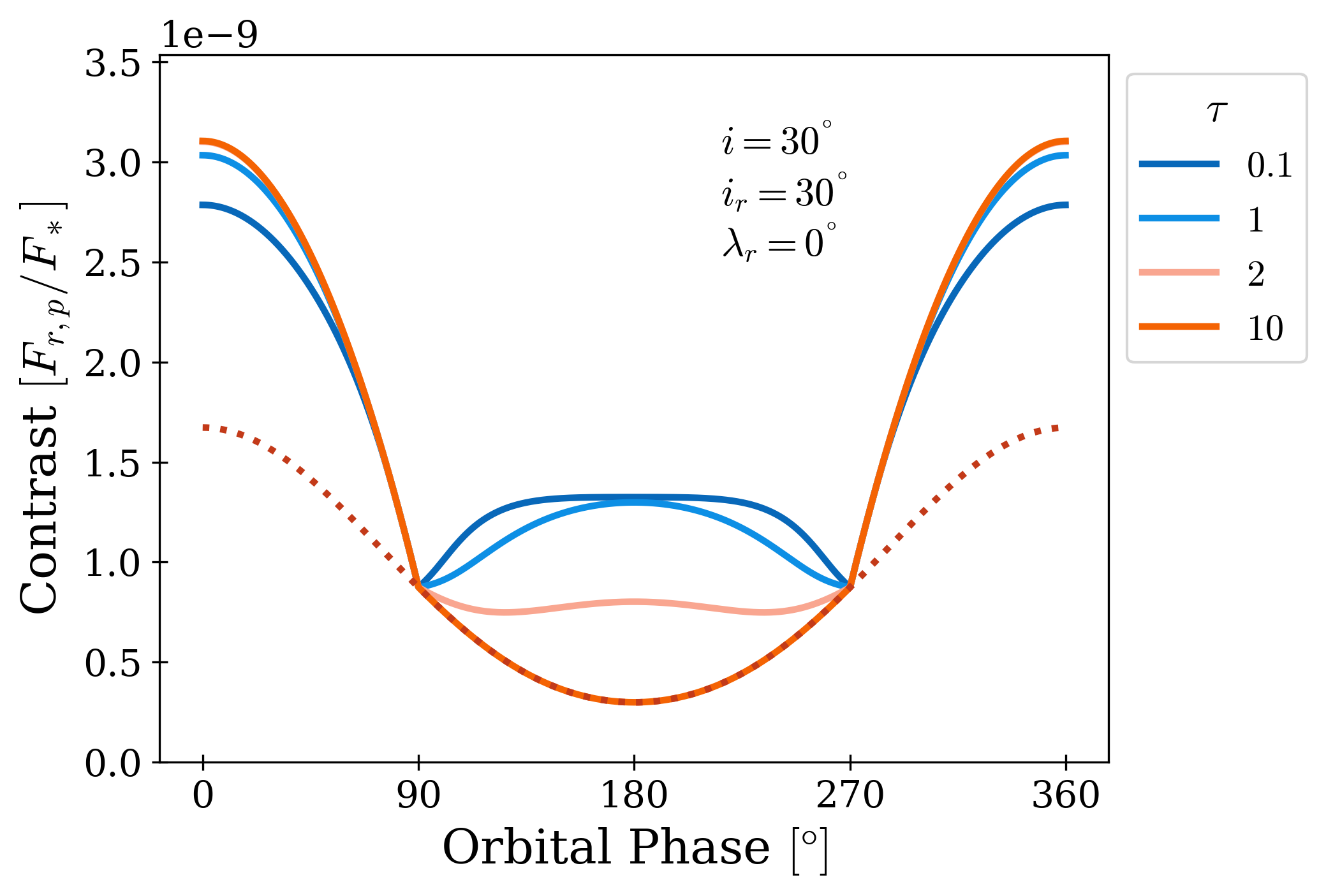}
    \caption{Same as Figure \ref{Obliquity}, showing changing normal optical thickness for constant orbit inclination, ring obliquity, and ring inclination longitude. {The dotted red line represents the ringless planet, and the solid lines represent the ringed planet phase curves. }When increasing the normal ring thickness, the transmitted portion of the ringed phase curve is reduced. When the ring is entirely opaque, the second reflected light maxima is lost.
}
\label{tau}
\end{figure}

    The ring-normal optical thickness $\tau$ primarily influences the transmitted component of the ringed phase curve, although it also has some impact on the reflected component. Shown in Figure \ref{tau}, as the optical thickness increases, less light is transmitted through the ring. For a fully opaque, optically thick ring, the transmission-driven second maximum in the ringed planet phase curve disappears. In this work, we only show examples of partially transmissive rings ($\tau=1$).  However, all known ring systems in the Solar System are at least partially transmissive, although their optical depths vary widely. As a result, the four-extrema signature explored here represents a plausible, though likely not universal, observational outcome for exoring systems. Because the four-extrema structure is the key feature that enables the identification of exorings in phase curves, optically thick rings are likely to be more difficult to detect. In this limit, the phase curve more closely resembles that of a higher-albedo or larger planet, and the ring signature may be difficult to distinguish {\citep{HaslerRings}}.

\section{Simulated CGI Observations of $\epsilon$~Eri~ {\MakeLowercase{b}}} \label{sec:epsErib}

\begin{figure*}
\centering
\includegraphics[width=0.65\textwidth]{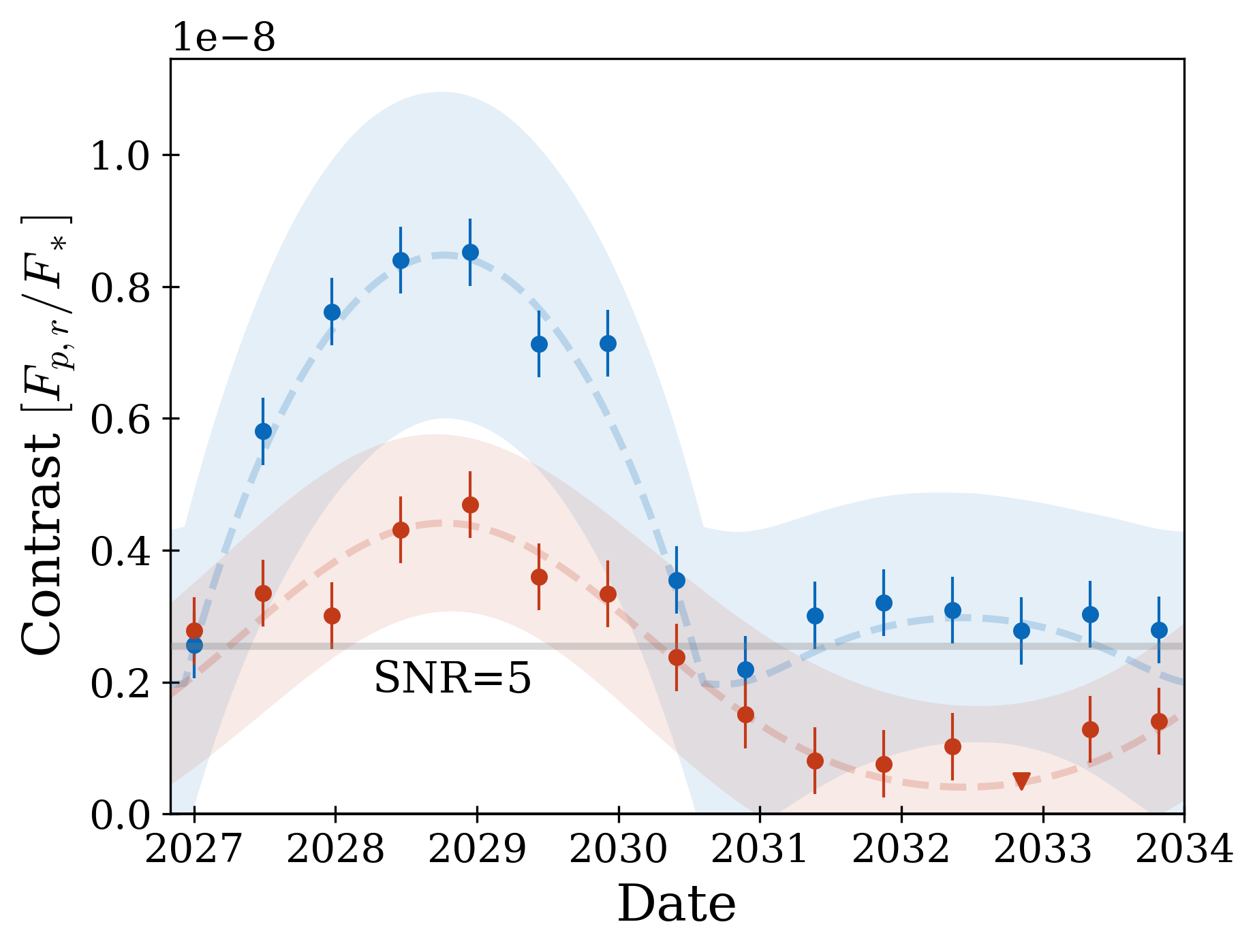}
    \caption{Simulated $\epsilon$~Eri~b data with (blue) and without (red) a ring compared to optimistic CGI detection limits (SNR=5){ and the continuous, modeled phase curve (dashed lines)}. {The observation dates are chosen to display $\epsilon$~Eri~b's full period, which extends past CGI's expected mission length. The uncertainties on simulated Roman contrasts are calculated from the \texttt{corgietc} Roman CGI exposure time calculator \citep{corgietc}, while the shaded regions represent the uncertainty on the orbit parameters reported in the literature for this planet \citep{2025AJ....170..301T}, shown in Table \ref{parameters}. To facilitate comparison between the ringed and ringless case, we artificially assume the central values in Table \ref{parameters} are known exactly.} With a ring, the maximum contrast is two times greater than that of the ringless case. Additionally, the ringed case demonstrates the two maxima per period. Both the unexpected increase in contrast and the four extrema pattern make the presence of a ring apparent. If the possible presence of a ring was ignored after making an observation, the increased contrast could mischaracterize the planet. Additionally, the potential contrast increase provided by a ring could expand CGI's target list to include objects expected to be below the contrast limit. Note that there are several epochs where the planet without a ring would not be detected (e.g., well below 5$\sigma$ limit).}
\label{epsErib}
\end{figure*}

    {Previous searches for $\epsilon$~Eridani~b ($\epsilon$~Eri~b) reached maximum contrast sensitivities of $3.0\times10^{-7}$, too low for the direct detection of the planet \citep{2019AJ....157...33M, 2021AJ....162..181L, 2024A&A...687A..74T, llopsayson2025searchingplanetsorbitingepsiloneridani, sanghi2026worldsdooriiiindirect}. Now, CGI will offer contrast sensitivities of $10^{-8}$ or better \citep{2023SPIE12680E..0TB, zellem2022nancygraceromanspace}, once again making $\epsilon$~Eri~b a favorable target for direct imaging with CGI, especially if the planet might have a ring.} We use our reflected light phase curve models to generate simulated Band 1 \citep[the primary CGI band with a central wavelength of 575~nm with a 10\% bandpass][]{zellem2022nancygraceromanspace} CGI observations of $\epsilon$~Eri~b with and without a Saturn-like ring, shown in Figure \ref{epsErib}. 
    
    The ringless case uses Equation~\ref{eq: Cahoy}, while the ringed case includes the ring contribution from Equation~\ref{eq: RingTotal}. We adopt the Saturn ring parameters and the measured $\epsilon$~Eri~b system parameters given in Table \ref{parameters}. {We assume we know the exact orbital parameters to retain the phase curve's distinctive features, but include the $1\sigma$ deviation about the continuous phase curve due to the uncertainty in the orbital parameters.} We then compute the contrast as a function of orbital phase for a ringed and ringless $\epsilon$~Eri~b over one orbital period{. The observation dates were chosen to span the full phase curve, but extend past the actual CGI mission dates (e.g., planned mission duration is 2026-2028, with possible extension to ~2031).} We select each specific date to fall within accessible observing windows determined using the \texttt{CGI\_target\_accessibility} tool \citep{target_accessibility}.

    We then use the \texttt{corgietc} Roman CGI exposure time calculator \citep{corgietc} to determine the achievable signal-to-noise ratios (SNRs) with ten-hour exposures. We use the tool's default value for the local zodiacal light (equivalent to a local zodiacal brightness of 23 $\frac{\text{mag}}{\text{arcsec}^{2}}$){, an exozodiacal light value of 296.6 zodis \citep{2020AJ....159..177E},} and the \texttt{optimistic observing mode} setting in Band 1.
    
    For each observation date, we iterate over an array of SNRs to identify the value corresponding to a ten-hour exposure time. We then compute the uncertainty by dividing the achieved contrast on a given date by the corresponding SNR. We add Gaussian noise to each simulated contrast measurement, with the standard deviation set by the corresponding contrast uncertainty derived from the CGI exposure-time calculation.

\begin{table}
\centering
\caption{Summary of parameters used for $\epsilon$~Eri~b and exoring reflected light phase curve. {Ring parameters are values for a Saturn-like ring.}} \label{parameters}
\medskip
\begin{tabular}{lccc}
\hline
Parameter & Value & Unit & Source$^{a}$ \\
\hline
\multicolumn{4}{l}{\textbf{Planet parameters}} \\
        $r_p$ & 1.0 &  R\textsubscript{Jup} & M21$^{b}$ \\
        $A_p$ & 0.3 & - & A04\\
        $T_p$ & {2458017.5$^{+1740.0}_{-447.0}$} & days & {T25}\\
        $\omega_p$ & {268$^{+62}_{-105}$} & deg & {T25}\\
        $i$ & {40$^{+6}_{-5}$} & deg & {T25}\\
        $P$ & {2680$\pm30$} & days & {T25}\\
        $a$ & {3.53$\pm0.04$} & AU & {T25}\\
\multicolumn{4}{l}{\textbf{Ring parameters}} \\
        $r_i$ & 1.5$r_p$ & R\textsubscript{Jup} & A04\\
        $r_o$ & 2.3$r_p$ & R\textsubscript{Jup} & A04\\
        $\omega_0$ & 0.7 & - & A04\\
        $\tau$ & 1.0 & - & A04\\
        $i_r$ & 30 & deg & A04\\
        $\lambda_r$ & 0 & deg & A04\\
\hline
\end{tabular}     \\ 
      $^{a}$References are M21: \cite{Marley2021}, A04: \cite{Arnold_2004} and {T25: \cite{2025AJ....170..301T}}.\\
      $^{b}$Sonora-Bobcat evolutionary models used to estimate $\epsilon$~Eri~b planetary radius.
      \vspace{2mm}
\end{table}

    As shown in Figure \ref{epsErib}, without a ring, the contrast of $\epsilon$~Eri~b only surpasses an optimistic SNR of five when the planet is near full phase. Near full phase, from {2027 to 2030}, the planet-only SNR does not exceed {10} in the 10-hour observations. From {2031 to 2034}, when the planet is near crescent phase, the ringless $\epsilon$~Eri~b's SNR dips below five and the contrast approaches zero. Therefore, for a large fraction of its period, the ringless $\epsilon$~Eri~b cannot be detected at statistically significant levels.
    
    With a Saturn-like ring, $\epsilon$~Eri~b appears up to two times brighter and displays the distinct four extrema signature. Additionally, the ringed $\epsilon$~Eri~b only falls below the optimistic detection limit when the ring is near edge-on to the star and not contributing to the total contrast. When the ring reflects the most light in {2028}, the SNR approaches {18}. When the ring is transmitting light from {2031 to 2034}, the SNR approaches seven. At most epochs, the ringed planet has a higher contrast than the ringless planet. However, when the ring is edge-on to the star in {2027 and 2031}, the ringed planet contrasts are similar to the ringless planet contrasts. Overall, these results show that rings can significantly enhance detectability with CGI and can produce observable signals even at phases where the planet alone would be faint.

\section{Discussion} \label{sec:discuss}

\subsection{Comparison with Previous Models and Future Work}

    The model presented here captures the first-order detectability of exorings with Roman, though it can be further refined and improved in future work. Our model does not include a scattering phase function or specific particle properties. More detailed modeling, such as ray tracing, could capture these effects, as well as mutual illumination and shadowing {\citep{HaslerRings}}, and specific ring composition models.{ For example, if the ring were not infinitely thin and contained finite particles, the ring may reflect and transmit light even when edge-on to the star. \citep{Veenstra_2025}. Ignoring this effect, our model underestimates the ring flux in edge-on cases.} While such models would alter the contrast and intricate structure of the phase curve, they would not eliminate the primary signatures or the underlying four-extrema structure identified here. Additionally, if phase curves are observed with Roman, further work will be needed to develop tools capable of fitting combined planet+ring phase-curve models.

    Understanding the impact of rings on CGI measurements is essential for accurate interpretation. For example, a Saturn-like ring around $\epsilon$~Eri~b could increase the observed flux by up to a factor of two (Figure \ref{epsErib}), potentially leading to the mischaracterization of the planet as larger or higher-albedo if the ring contribution is not accounted for. Disentangling these effects may require multi-epoch observations, though obtaining dense phase coverage for all targets is likely impractical. More broadly, rings introduce degeneracies in inferred planetary properties, including radius and albedo, and can bias retrievals if not properly modeled. While phase-curve analyses offer a pathway to break these degeneracies for individual systems, constraining the occurrence rate of exorings will be critical for interpreting populations and informing future direct imaging missions such as the Habitable Worlds Observatory \citep{2021pdaa.book.....N}.

    In future work, we will build a phase curve fitter that can identify a ring signature in multi-epoch CGI observations. We plan to use this fitter to quantify the phase coverage needed to distinguish ringed from ringless phase curves as a function of planet and ring parameters.

\subsection{Implications for Detectability with Roman CGI}

\begin{deluxetable*}{lcccccc}
\tablewidth{0pt}
\tablecaption{Example Target List and Estimated Ring Detectability  \label{tab:targets}}
\tablehead{
\colhead{Name} & 
\colhead{Host $V$ mag} & 
\colhead{Planet Contrast$^a$} &
\colhead{Ring {+Planet }Contrast$^a$} & 
\colhead{Separation$^b$ (")} & 
\colhead{$M_p$ ($M_J$)} & 
\colhead{$a$ (AU)}
}
\startdata
HD 154345 b & 6.76 & {$2\times10^{-9}$} & {$6\times10^{-9}$} & 0.23 & $1.2\pm0.1$ & 4.16 \\
$\pi$ Men b & 5.65 & {$2\times10^{-9}$}& {$7\times10^{-9}$} & 0.18 & $12.3\pm1$ & 3.31 \\
55 Cnc d & 5.95 & {$1\times10^{-9}$}& {$3\times10^{-9}$} & 0.43 & $3.8\pm0.1$ & 5.45 \\
HD 114613 b & 4.85 & {$1\times10^{-9}$}& {$4\times10^{-9}$} & 0.26 & $0.36\pm0.03$ & 5.34 \\
$\upsilon$ And d & 4.09 & {$5\times10^{-9}$}& {$1\times10^{-8}$} & 0.19 & $10\pm2$ & 2.55 \\
HD 217107 c & 6.16 & {$9\times10^{-10}$}& {$2\times10^{-9}$}& 0.30 & $4.4\pm0.1$ & 5.92 \\
14 Her b & 6.67 & {$4\times10^{-9}$}& {$1\times10^{-8}$} & 0.15 & $8.9\pm1.5$ & 2.84 \\
47 UMa c & 5.10 & {$3\times10^{-9}$}& {$8\times10^{-9}$} & 0.26 & $0.5\pm0.1$ & 3.6 \\
47 UMa b & 5.10 & {$7\times10^{-9}$}& {$2\times10^{-8}$} & 0.15 & $2.5\pm0.1$ & 2.1 \\
HD 190360 b & 5.71 & {$2\times10^{-9}$}& {$6\times10^{-9}$} & 0.25 & $1.5\pm0.2$ & 3.92 \\
$\psi^1$ Dra B b & 5.7 & {$2\times10^{-9}$}& {$5\times10^{-9}$} & 0.20 & $1.5\pm0.1$ & 4.43 \\
HD 219077 b & 6.12 & {$6\times10^{-10}$}& {$2\times10^{-9}$} & 0.24 & $13.4\pm0.8$ & 7.03 \\
HD 134987 c & 6.45 & {$1\times10^{-9}$} & {$3\times10^{-9}$} & 0.26 & $0.82\pm0.03$ & 5.8 \\
$\mu$ Ara c & 5.15 & {$1\times10^{-9}$}& {$3\times10^{-9}$} & 0.33 & $\sim1.8$ & 5.24 \\
HD 219134 h & 5.57 & {$4\times10^{-9}$}& {$1\times10^{-8}$} & 0.47 & $0.31\pm0.01$ & 3.06 \\
HD 142 c & 7.96 & {$8\times10^{-9}$} & {$2\times10^{-8}$} & 0.25 & $\sim0.89$ & 1.96 \\
$\gamma$ Cep b & 3.22 & {$7\times10^{-9}$} & {$2\times10^{-8}$} & 0.16 & $1.7\pm0.1$ & 2.15 \\
HR 5183 b & 6.3 & {$9\times10^{-11}$} & {$3\times10^{-10}$} & 0.57 & $3.2\pm0.1$ & 18.0 \\
$\tau$ Cet e & 3.5 & {$2\times10^{-9}$} & {$7\times10^{-9}$} & 0.15 & $0.01\pm0.003$ & 0.54 \\
HD 62509 b & 1.15 & {$1\times10^{-8}$} & {$3\times10^{-8}$} & 0.16 & $2.9\pm0.1$ & 1.69 \\
$\epsilon$ Eri b & 3.73 & {$3\times10^{-9}$} & {$8\times10^{-9}$} & 1.1 & $0.98\pm0.1$ & 3.53 \\
HD 192310 c & 6.13 & {$5\times10^{-9}$} & {$1\times10^{-8}$} & 0.13 & $0.08\pm0.02$ & 1.18 \\
HD 30562 b & 5.77 & {$6\times10^{-9}$} & {$2\times10^{-8}$} & 0.10 & $1.29\pm0.1$ & 2.3 \\
$\tau$ Cet f & 3.5 & {$4\times10^{-10}$} & {$1\times10^{-9}$} & 0.37 & $0.01\pm0.004$ & 1.33 \\
\enddata
\tablenotetext{a}{{Optimistic contrasts estimated with Saturn-like ring parameters (Table \ref{parameters}) and systems viewed face-on ($i=0^\circ$). The ring obliquity $i_r$ is $30^\circ$ and the ring inclination longitude $\lambda_r$ is $0^\circ$. The planets are in quadrature, and the rings are observed when reflected light is maximized.}}
\tablenotetext{b}{Projected separation assuming $a/d$; edge-on systems may fall within the inner working angle.}
\end{deluxetable*}

    As a crude but illustrative estimate of the impact of exorings on detectability, we consider how the presence of Saturn-like ring systems would alter the observable contrast of known radial velocity giant planets observed with CGI. In particular, we ask how many systems that are nominally below the detection threshold for planetary reflected light could instead become detectable due to the additional flux contributed by a ring system.

    To explore this, we adopt a simple approximation in which the reflected light signal from the system is dominated by the projected area of the ring for a Saturn-like ring system viewed in the most optimistic (face-on) configuration. We apply this approximation to a sample of known nearby giant exoplanets \citep[Table~\ref{tab:targets};][]{Carri_n_Gonz_lez_2021}, computing the contrast required to detect a Saturn-sized ring system at the projected separation of each planet. Notably, many of the systems in this sample would not be considered viable targets for direct imaging of the planet alone, but become accessible once the contribution from rings is included.

    The results illustrate that although a significant fraction of known radial velocity planets lie near or just below the nominal contrast limits of CGI, with the addition of a Saturn-like ring system, many of these objects shift into the detectable regime at contrasts of $\sim 10^{-8}$. This implies that Roman may be sensitive to a population of ringed planets that would otherwise remain undetected in reflected light surveys.

    From a statistical perspective, this has important implications for constraining the occurrence rate of exorings. If the occurrence rate of Saturn-like rings matches that of Solar System giant planets (i.e., $\sim$25\%), then a survey of $\sim$20 targets would be expected to yield $\sim$5 detections. Conversely, a null result would place meaningful upper limits on the prevalence of such systems, ruling out Solar System-like occurrence rates at the $\sim$2--3$\sigma$ level. Even with this highly simplified treatment, it is clear that both detections and non-detections provide strong constraints on exoring populations.

    We emphasize that this calculation is intentionally approximate. It neglects the effects of ring orientation, optical depth variations, shadowing, and scattering phase functions, all of which can significantly modify the observed flux. In particular, the detectability of rings depends sensitively on inclination and viewing geometry. Nevertheless, this first-order estimate demonstrates that rings can substantially enhance the reflected-light signal and expand the pool of observable targets. Some of the planets listed in Table~\ref{tab:targets} are terrestrial. While large ring systems may be less likely around terrestrial planets, this remains observationally unconstrained; measuring the occurrence rate across different planet populations, even in the absence of detections, is therefore important.

\section{Conclusion} \label{sec:conclusion}

    In this work, we investigate how exorings can affect Roman CGI direct imaging observations using an analytic ring model, focusing on reflected light phase curves and the impact of ring parameters on phase curve geometry. We develop a model for ringed planet phase curves, emphasizing optically thin rings, simulate multi-epoch observations of ringed and ringless $\epsilon$~Eri~b, and explore how viewing geometry influences the resulting signals. Our main findings are:

\begin{itemize}
    \item Rings can significantly enhance a planet’s reflected light signal, in some cases increasing the observed flux by factors of several to hundreds and bringing otherwise undetectable targets within CGI’s sensitivity.
    \item Optically thin rings (similar to Saturn's) produce a distinctive four-extrema phase-curve signature, providing a clear and testable method for identifying and confirming exorings in multi-epoch observations.
    \item For $\epsilon$~Eri~b, a Saturn-like ring would enable detection at $\gtrsim 5\sigma$ across nearly all orbital phases, whereas the ringless planet is often below the detection threshold.
    \item Ringed planets are generally easier to detect than their ringless counterparts, potentially expanding the number of known RV exoplanet targets accessible to CGI if Saturn-like rings are common.
    \item Rings introduce degeneracies in inferred planetary properties, including radius and albedo; if unaccounted for, planets may be mischaracterized as larger or more reflective than they are. Phase-curve observations of optically thin rings provide a practical pathway to breaking these degeneracies, while optically thick rings will be more difficult to distinguish from unringed planets.
    \item CGI observations are well positioned to place the first meaningful constraints on the occurrence rate of Saturn-like exorings.
    \item Further work is needed to refine ring models and develop tools for fitting combined planet+ring phase curves, enabling robust interpretation of future CGI data.
\end{itemize}

\vspace{2mm}

\section*{Acknowledgments}

    {We thank an anonymous referee for their constructive comments, which improved the quality of this manuscript.} We thank {Vanessa Bailey for her feedback on our analysis, and} Professor Michael Meyer for his guidance and feedback in the University of Michigan's \textit{Scientific Writing and Communication in Astronomy} course. Additionally, Lane thanks their classmates for their thoughtful reviews of the early drafts of this manuscript. This research has made use of the NASA Exoplanet Archive, which is operated by the California Institute of Technology, under contract with the National Aeronautics and Space Administration under the Exoplanet Exploration Program.

\facilities{Roman Space Telescope}

\software{{\tt Astropy} \citep{2013A&A...558A..33A, 2018AJ....156..123A, 2022ApJ...935..167A}, {\tt Blender} \citep{Blender}, {\tt CGI\_target\_accessibility} \citep{target_accessibility}, {\tt corgietc} \citep{corgietc}, {\tt NumPy} \citep{harris2020array}, {\tt Matplotlib} \citep{Hunter:2007}}

\bibliography{main.bib}{}

\begin{thebibliography}{}
\expandafter\ifx\csname natexlab\endcsname\relax\def\natexlab#1{#1}\fi
\providecommand{\url}[1]{\href{#1}{#1}}
\providecommand{\dodoi}[1]{doi:~\href{http://doi.org/#1}{\nolinkurl{#1}}}
\providecommand{\doeprint}[1]{\href{http://ascl.net/#1}{\nolinkurl{http://ascl.net/#1}}}
\providecommand{\doarXiv}[1]{\href{https://arxiv.org/abs/#1}{\nolinkurl{https://arxiv.org/abs/#1}}}

\bibitem[{{Aizawa} {et~al.}(2017){Aizawa}, {Uehara}, {Masuda}, {Kawahara}, \& {Suto}}]{2017AJ....153..193A}
{Aizawa}, M., {Uehara}, S., {Masuda}, K., {Kawahara}, H., \& {Suto}, Y. 2017, \aj, 153, 193, \dodoi{10.3847/1538-3881/aa6336}

\bibitem[{{Akeson} {et~al.}(2019){Akeson}, {Armus}, {Bachelet}, {Bailey}, {Bartusek}, {Bellini}, {Benford}, {Bennett}, {Bhattacharya}, {Bohlin}, {Boyer}, {Bozza}, {Bryden}, {Calchi Novati}, {Carpenter}, {Casertano}, {Choi}, {Content}, {Dayal}, {Dressler}, {Dor{\'e}}, {Fall}, {Fan}, {Fang}, {Filippenko}, {Finkelstein}, {Foley}, {Furlanetto}, {Kalirai}, {Gaudi}, {Gilbert}, {Girard}, {Grady}, {Greene}, {Guhathakurta}, {Heinrich}, {Hemmati}, {Hendel}, {Henderson}, {Henning}, {Hirata}, {Ho}, {Huff}, {Hutter}, {Jansen}, {Jha}, {Johnson}, {Jones}, {Kasdin}, {Kelly}, {Kirshner}, {Koekemoer}, {Kruk}, {Lewis}, {Macintosh}, {Madau}, {Malhotra}, {Mandel}, {Massara}, {Masters}, {McEnery}, {McQuinn}, {Melchior}, {Melton}, {Mennesson}, {Peeples}, {Penny}, {Perlmutter}, {Pisani}, {Plazas}, {Poleski}, {Postman}, {Ranc}, {Rauscher}, {Rest}, {Roberge}, {Robertson}, {Rodney}, {Rhoads}, {Rhodes}, {Ryan}, {Sahu}, {Sand}, {Scolnic}, {Seth}, {Shvartzvald}, {Siellez}, {Smith}, {Spergel}, {Stassun}, {Street}, {Strolger}, {Szalay},
  {Trauger}, {Troxel}, {Turnbull}, {van der Marel}, {von der Linden}, {Wang}, {Weinberg}, {Williams}, {Windhorst}, {Wollack}, {Wu}, {Yee}, \& {Zimmerman}}]{2019arXiv190205569A}
{Akeson}, R., {Armus}, L., {Bachelet}, E., {et~al.} 2019, arXiv e-prints, arXiv:1902.05569, \dodoi{10.48550/arXiv.1902.05569}

\bibitem[{Akinsanmi {et~al.}(2020)Akinsanmi, Santos, Faria, Oshagh, Barros, Santerne, \& Charnoz}]{Akinsanmi_2020}
Akinsanmi, B., Santos, N.~C., Faria, J.~P., {et~al.} 2020, Astronomy \& Astrophysics, 635, L8, \dodoi{10.1051/0004-6361/202037618}

\bibitem[{Arnold \& Schneider(2004)}]{Arnold_2004}
Arnold, L., \& Schneider, J. 2004, Astronomy \& Astrophysics, 420, 1153–1162, \dodoi{10.1051/0004-6361:20035720}

\bibitem[{{Astropy Collaboration} {et~al.}(2013){Astropy Collaboration}, {Robitaille}, {Tollerud}, {Greenfield}, {Droettboom}, {Bray}, {Aldcroft}, {Davis}, {Ginsburg}, {Price-Whelan}, {Kerzendorf}, {Conley}, {Crighton}, {Barbary}, {Muna}, {Ferguson}, {Grollier}, {Parikh}, {Nair}, {Unther}, {Deil}, {Woillez}, {Conseil}, {Kramer}, {Turner}, {Singer}, {Fox}, {Weaver}, {Zabalza}, {Edwards}, {Azalee Bostroem}, {Burke}, {Casey}, {Crawford}, {Dencheva}, {Ely}, {Jenness}, {Labrie}, {Lim}, {Pierfederici}, {Pontzen}, {Ptak}, {Refsdal}, {Servillat}, \& {Streicher}}]{2013A&A...558A..33A}
{Astropy Collaboration}, {Robitaille}, T.~P., {Tollerud}, E.~J., {et~al.} 2013, \aap, 558, A33, \dodoi{10.1051/0004-6361/201322068}

\bibitem[{{Astropy Collaboration} {et~al.}(2018){Astropy Collaboration}, {Price-Whelan}, {Sip{\H{o}}cz}, {G{\"u}nther}, {Lim}, {Crawford}, {Conseil}, {Shupe}, {Craig}, {Dencheva}, {Ginsburg}, {VanderPlas}, {Bradley}, {P{\'e}rez-Su{\'a}rez}, {de Val-Borro}, {Aldcroft}, {Cruz}, {Robitaille}, {Tollerud}, {Ardelean}, {Babej}, {Bach}, {Bachetti}, {Bakanov}, {Bamford}, {Barentsen}, {Barmby}, {Baumbach}, {Berry}, {Biscani}, {Boquien}, {Bostroem}, {Bouma}, {Brammer}, {Bray}, {Breytenbach}, {Buddelmeijer}, {Burke}, {Calderone}, {Cano Rodr{\'\i}guez}, {Cara}, {Cardoso}, {Cheedella}, {Copin}, {Corrales}, {Crichton}, {D'Avella}, {Deil}, {Depagne}, {Dietrich}, {Donath}, {Droettboom}, {Earl}, {Erben}, {Fabbro}, {Ferreira}, {Finethy}, {Fox}, {Garrison}, {Gibbons}, {Goldstein}, {Gommers}, {Greco}, {Greenfield}, {Groener}, {Grollier}, {Hagen}, {Hirst}, {Homeier}, {Horton}, {Hosseinzadeh}, {Hu}, {Hunkeler}, {Ivezi{\'c}}, {Jain}, {Jenness}, {Kanarek}, {Kendrew}, {Kern}, {Kerzendorf}, {Khvalko}, {King}, {Kirkby}, {Kulkarni},
  {Kumar}, {Lee}, {Lenz}, {Littlefair}, {Ma}, {Macleod}, {Mastropietro}, {McCully}, {Montagnac}, {Morris}, {Mueller}, {Mumford}, {Muna}, {Murphy}, {Nelson}, {Nguyen}, {Ninan}, {N{\"o}the}, {Ogaz}, {Oh}, {Parejko}, {Parley}, {Pascual}, {Patil}, {Patil}, {Plunkett}, {Prochaska}, {Rastogi}, {Reddy Janga}, {Sabater}, {Sakurikar}, {Seifert}, {Sherbert}, {Sherwood-Taylor}, {Shih}, {Sick}, {Silbiger}, {Singanamalla}, {Singer}, {Sladen}, {Sooley}, {Sornarajah}, {Streicher}, {Teuben}, {Thomas}, {Tremblay}, {Turner}, {Terr{\'o}n}, {van Kerkwijk}, {de la Vega}, {Watkins}, {Weaver}, {Whitmore}, {Woillez}, {Zabalza}, \& {Astropy Contributors}}]{2018AJ....156..123A}
{Astropy Collaboration}, {Price-Whelan}, A.~M., {Sip{\H{o}}cz}, B.~M., {et~al.} 2018, \aj, 156, 123, \dodoi{10.3847/1538-3881/aabc4f}

\bibitem[{{Astropy Collaboration} {et~al.}(2022){Astropy Collaboration}, {Price-Whelan}, {Lim}, {Earl}, {Starkman}, {Bradley}, {Shupe}, {Patil}, {Corrales}, {Brasseur}, {N{\"o}the}, {Donath}, {Tollerud}, {Morris}, {Ginsburg}, {Vaher}, {Weaver}, {Tocknell}, {Jamieson}, {van Kerkwijk}, {Robitaille}, {Merry}, {Bachetti}, {G{\"u}nther}, {Aldcroft}, {Alvarado-Montes}, {Archibald}, {B{\'o}di}, {Bapat}, {Barentsen}, {Baz{\'a}n}, {Biswas}, {Boquien}, {Burke}, {Cara}, {Cara}, {Conroy}, {Conseil}, {Craig}, {Cross}, {Cruz}, {D'Eugenio}, {Dencheva}, {Devillepoix}, {Dietrich}, {Eigenbrot}, {Erben}, {Ferreira}, {Foreman-Mackey}, {Fox}, {Freij}, {Garg}, {Geda}, {Glattly}, {Gondhalekar}, {Gordon}, {Grant}, {Greenfield}, {Groener}, {Guest}, {Gurovich}, {Handberg}, {Hart}, {Hatfield-Dodds}, {Homeier}, {Hosseinzadeh}, {Jenness}, {Jones}, {Joseph}, {Kalmbach}, {Karamehmetoglu}, {Ka{\l}uszy{\'n}ski}, {Kelley}, {Kern}, {Kerzendorf}, {Koch}, {Kulumani}, {Lee}, {Ly}, {Ma}, {MacBride}, {Maljaars}, {Muna}, {Murphy}, {Norman},
  {O'Steen}, {Oman}, {Pacifici}, {Pascual}, {Pascual-Granado}, {Patil}, {Perren}, {Pickering}, {Rastogi}, {Roulston}, {Ryan}, {Rykoff}, {Sabater}, {Sakurikar}, {Salgado}, {Sanghi}, {Saunders}, {Savchenko}, {Schwardt}, {Seifert-Eckert}, {Shih}, {Jain}, {Shukla}, {Sick}, {Simpson}, {Singanamalla}, {Singer}, {Singhal}, {Sinha}, {Sip{\H{o}}cz}, {Spitler}, {Stansby}, {Streicher}, {{\v{S}}umak}, {Swinbank}, {Taranu}, {Tewary}, {Tremblay}, {de Val-Borro}, {Van Kooten}, {Vasovi{\'c}}, {Verma}, {de Miranda Cardoso}, {Williams}, {Wilson}, {Winkel}, {Wood-Vasey}, {Xue}, {Yoachim}, {Zhang}, {Zonca}, \& {Astropy Project Contributors}}]{2022ApJ...935..167A}
{Astropy Collaboration}, {Price-Whelan}, A.~M., {Lim}, P.~L., {et~al.} 2022, \apj, 935, 167, \dodoi{10.3847/1538-4357/ac7c74}

\bibitem[{Bailey(2021)}]{target_accessibility}
Bailey, V. 2021, {CGI\_target\_accessibility},  Github.
\newblock \url{https://github.com/nasavbailey/CGI_target_accessibility}

\bibitem[{{Bailey} {et~al.}(2023){Bailey}, {Bendek}, {Monacelli}, {Baker}, {Bedrosian}, {Cady}, {Douglas}, {Groff}, {Hildebrandt}, {Kasdin}, {Krist}, {Macintosh}, {Mennesson}, {Morrissey}, {Poberezhskiy}, {Subedi}, {Rhodes}, {Roberge}, {Ygouf}, {Zellem}, {Zhao}, \& {Zimmerman}}]{2023SPIE12680E..0TB}
{Bailey}, V.~P., {Bendek}, E., {Monacelli}, B., {et~al.} 2023, in Society of Photo-Optical Instrumentation Engineers (SPIE) Conference Series, Vol. 12680, Society of Photo-Optical Instrumentation Engineers (SPIE) Conference Series, 126800T, \dodoi{10.1117/12.2679036}

\bibitem[{Barnes \& Fortney(2004)}]{Barnes_2004}
Barnes, J.~W., \& Fortney, J.~J. 2004, The Astrophysical Journal, 616, 1193, \dodoi{10.1086/425067}

\bibitem[{{Barnes} \& {Fortney}(2004)}]{2004ApJ...616.1193B}
{Barnes}, J.~W., \& {Fortney}, J.~J. 2004, \apj, 616, 1193, \dodoi{10.1086/425067}

\bibitem[{Benisty {et~al.}(2021)Benisty, Bae, Facchini, Keppler, Teague, Isella, Kurtovic, Pérez, Sierra, Andrews, Carpenter, Czekala, Dominik, Henning, Menard, Pinilla, \& Zurlo}]{Benisty_2021}
Benisty, M., Bae, J., Facchini, S., {et~al.} 2021, The Astrophysical Journal Letters, 916, L2, \dodoi{10.3847/2041-8213/ac0f83}

\bibitem[{{Blender Online Community}(2018)}]{Blender}
{Blender Online Community}. 2018, Blender - a 3D modelling and rendering package, Blender Foundation, Stichting Blender Foundation, Amsterdam.
\newblock \url{http://www.blender.org}

\bibitem[{Bottke {et~al.}(2005)Bottke, Durda, Nesvorný, Jedicke, Morbidelli, Vokrouhlický, \& Levison}]{BOTTKE2005111}
Bottke, W.~F., Durda, D.~D., Nesvorný, D., {et~al.} 2005, Icarus, 175, 111, \dodoi{https://doi.org/10.1016/j.icarus.2004.10.026}

\bibitem[{Bowens-Rubin {et~al.}(2025)Bowens-Rubin, Limbach, Hopper, Stephenson, Garza, Fletcher, \& Hedman}]{bowensrubin2025detectionexoringsreflectedlight}
Bowens-Rubin, R., Limbach, M.~A., Hopper, S., {et~al.} 2025, On the Detection of Exorings in Reflected Light with JWST NIRCam.
\newblock \doarXiv{2509.07118}

\bibitem[{Cahoy {et~al.}(2010)Cahoy, Marley, \& Fortney}]{Cahoy_2010}
Cahoy, K.~L., Marley, M.~S., \& Fortney, J.~J. 2010, The Astrophysical Journal, 724, 189–214, \dodoi{10.1088/0004-637x/724/1/189}

\bibitem[{Canup(2010)}]{article}
Canup, R. 2010, Nature, 468, 943, \dodoi{10.1038/nature09661}

\bibitem[{Carrión-González {et~al.}(2021)Carrión-González, García~Muñoz, Santos, Cabrera, Csizmadia, \& Rauer}]{Carri_n_Gonz_lez_2021}
Carrión-González, O., García~Muñoz, A., Santos, N.~C., {et~al.} 2021, Astronomy \& Astrophysics, 651, A7, \dodoi{10.1051/0004-6361/202039993}

\bibitem[{{Cassese} {et~al.}(2026){Cassese}, {Lu}, {Rice}, {Yahalomi}, {Hattori}, {Vissapragada}, \& {Carter}}]{2026jwst.prop.11253}
{Cassese}, B., {Lu}, T., {Rice}, M., {et~al.} 2026, {luffy or Flashy? Resolving the Mystery of the Coldest-Known Super Puff}, JWST Proposal. Cycle 5, ID. \#11253

\bibitem[{{Charnoz} {et~al.}(2018){Charnoz}, {Canup}, {Crida}, \& {Dones}}]{2018prs..book..517C}
{Charnoz}, S., {Canup}, R.~M., {Crida}, A., \& {Dones}, L. 2018, in Planetary Ring Systems. Properties, Structure, and Evolution, ed. M.~S. {Tiscareno} \& C.~D. {Murray}, 517--538, \dodoi{10.1017/9781316286791.018}

\bibitem[{{Crida} {et~al.}(2025){Crida}, {Estrada}, {Nicholson}, \& {Murray}}]{2025SSRv..221...66C}
{Crida}, A., {Estrada}, P.~R., {Nicholson}, P.~D., \& {Murray}, C.~D. 2025, \ssr, 221, 66, \dodoi{10.1007/s11214-025-01189-z}

\bibitem[{{Elliot} {et~al.}(1977){Elliot}, {Dunham}, \& {Mink}}]{1977Natur.267..328E}
{Elliot}, J.~L., {Dunham}, E., \& {Mink}, D. 1977, \nat, 267, 328, \dodoi{10.1038/267328a0}

\bibitem[{{Ertel} {et~al.}(2020){Ertel}, {Defr{\`e}re}, {Hinz}, {Mennesson}, {Kennedy}, {Danchi}, {Gelino}, {Hill}, {Hoffmann}, {Mazoyer}, {Rieke}, {Shannon}, {Stapelfeldt}, {Spalding}, {Stone}, {Vaz}, {Weinberger}, {Willems}, {Absil}, {Arbo}, {Bailey}, {Beichman}, {Bryden}, {Downey}, {Durney}, {Esposito}, {Gaspar}, {Grenz}, {Haniff}, {Leisenring}, {Marion}, {McMahon}, {Millan-Gabet}, {Montoya}, {Morzinski}, {Perera}, {Pinna}, {Pott}, {Power}, {Puglisi}, {Roberge}, {Serabyn}, {Skemer}, {Su}, {Vaitheeswaran}, \& {Wyatt}}]{2020AJ....159..177E}
{Ertel}, S., {Defr{\`e}re}, D., {Hinz}, P., {et~al.} 2020, \aj, 159, 177, \dodoi{10.3847/1538-3881/ab7817}

\bibitem[{{Fortney}(2024)}]{2024arXiv240504587F}
{Fortney}, J.~J. 2024, arXiv e-prints, arXiv:2405.04587, \dodoi{10.48550/arXiv.2405.04587}

\bibitem[{Harris {et~al.}(2020)Harris, Millman, van~der Walt, Gommers, Virtanen, Cournapeau, Wieser, Taylor, Berg, Smith, Kern, Picus, Hoyer, van Kerkwijk, Brett, Haldane, del R{\'{i}}o, Wiebe, Peterson, G{\'{e}}rard-Marchant, Sheppard, Reddy, Weckesser, Abbasi, Gohlke, \& Oliphant}]{harris2020array}
Harris, C.~R., Millman, K.~J., van~der Walt, S.~J., {et~al.} 2020, Nature, 585, 357, \dodoi{10.1038/s41586-020-2649-2}

\bibitem[{{Hasler} {et~al.}(2026){Hasler}, {Greenbaum}, {Bryden}, {Bailey}, {Llop-Sayson}, {Lane}, {Limbach}, {Pearce}, {Ingalls}, \& {Lowrance}}]{HaslerRings}
{Hasler}, S.~N., {Greenbaum}, A.~Z., {Bryden}, G., {et~al.} 2026, \apj, \dodoi{10.3847/1538-4357/aeb012}

\bibitem[{{Hillier} {et~al.}(2007){Hillier}, {Green}, {McBride}, {Schwanethal}, {Postberg}, {Srama}, {Kempf}, {Moragas-Klostermeyer}, {McDonnell}, \& {Gr{\"u}n}}]{2007MNRAS.377.1588H}
{Hillier}, J.~K., {Green}, S.~F., {McBride}, N., {et~al.} 2007, \mnras, 377, 1588, \dodoi{10.1111/j.1365-2966.2007.11710.x}

\bibitem[{Hoch {et~al.}(2025)Hoch, Rowland, Petrus, Nasedkin, Ingebretsen, Kammerer, Perrin, D’Orazi, Balmer, Barman, Bonnefoy, Chauvin, Chen, De~Rosa, Girard, Gonzales, Kenworthy, Konopacky, Macintosh, Moran, Morley, Palma-Bifani, Pueyo, Ren, Rickman, Ruffio, Theissen, Ward-Duong, \& Zhang}]{Hoch_2025}
Hoch, K. K.~W., Rowland, M., Petrus, S., {et~al.} 2025, Nature, 643, 938–942, \dodoi{10.1038/s41586-025-09174-w}

\bibitem[{{Hubbard} {et~al.}(1986){Hubbard}, {Brahic}, {Sicardy}, {Elicer}, {Roques}, \& {Vilas}}]{1986Natur.319..636H}
{Hubbard}, W.~B., {Brahic}, A., {Sicardy}, B., {et~al.} 1986, \nat, 319, 636, \dodoi{10.1038/319636a0}

\bibitem[{Hunter(2007)}]{Hunter:2007}
Hunter, J.~D. 2007, Computing in Science \& Engineering, 9, 90, \dodoi{10.1109/MCSE.2007.55}

\bibitem[{{Hyodo} \& {Ohtsuki}(2015)}]{2015NatGe...8..686H}
{Hyodo}, R., \& {Ohtsuki}, K. 2015, Nature Geoscience, 8, 686, \dodoi{10.1038/ngeo2508}

\bibitem[{{Kasdin} {et~al.}(2020){Kasdin}, {Bailey}, {Mennesson}, {Zellem}, {Ygouf}, {Rhodes}, {Luchik}, {Zhao}, {Riggs}, {Seo}, {Krist}, {Kern}, {Tang}, {Nemati}, {Groff}, {Zimmerman}, {Macintosh}, {Turnbull}, {Debes}, {Douglas}, \& {Lupu}}]{2020SPIE11443E..1UK}
{Kasdin}, N.~J., {Bailey}, V.~P., {Mennesson}, B., {et~al.} 2020, in Society of Photo-Optical Instrumentation Engineers (SPIE) Conference Series, Vol. 11443, Space Telescopes and Instrumentation 2020: Optical, Infrared, and Millimeter Wave, ed. M.~{Lystrup} \& M.~D. {Perrin}, 114431U, \dodoi{10.1117/12.2562997}

\bibitem[{{Kempton} \& {Knutson}(2024)}]{2024RvMG...90..411K}
{Kempton}, E. M.-R., \& {Knutson}, H.~A. 2024, Reviews in Mineralogy and Geochemistry, 90, 411, \dodoi{10.2138/rmg.2024.90.12}

\bibitem[{Kenworthy \& Mamajek(2015)}]{Kenworthy_2015}
Kenworthy, M.~A., \& Mamajek, E.~E. 2015, The Astrophysical Journal, 800, 126, \dodoi{10.1088/0004-637x/800/2/126}

\bibitem[{Lacy {et~al.}(2019)Lacy, Shlivko, \& Burrows}]{Lacy_2019}
Lacy, B., Shlivko, D., \& Burrows, A. 2019, The Astronomical Journal, 157, 132, \dodoi{10.3847/1538-3881/ab0415}

\bibitem[{Liang {et~al.}(2021)Liang, Robnik, \& Seljak}]{Liang_2021}
Liang, Y., Robnik, J., \& Seljak, U. 2021, The Astronomical Journal, 161, 202, \dodoi{10.3847/1538-3881/abe6a7}

\bibitem[{Lietzow \& Wolf(2023)}]{Lietzow_2023}
Lietzow, M., \& Wolf, S. 2023, Astronomy \& Astrophysics, 671, A113, \dodoi{10.1051/0004-6361/202245474}

\bibitem[{{Llop-Sayson} {et~al.}(2021){Llop-Sayson}, {Wang}, {Ruffio}, {Mawet}, {Blunt}, {Absil}, {Bond}, {Brinkman}, {Bowler}, {Bottom}, {Chontos}, {Dalba}, {Fulton}, {Giacalone}, {Hill}, {Hirsch}, {Howard}, {Isaacson}, {Karlsson}, {Lubin}, {Madurowicz}, {Matthews}, {Morris}, {Perrin}, {Ren}, {Rice}, {Rosenthal}, {Ruane}, {Rubenzahl}, {Sun}, {Wallack}, {Xuan}, \& {Ygouf}}]{2021AJ....162..181L}
{Llop-Sayson}, J., {Wang}, J.~J., {Ruffio}, J.-B., {et~al.} 2021, \aj, 162, 181, \dodoi{10.3847/1538-3881/ac134a}

\bibitem[{Llop-Sayson {et~al.}(2025)Llop-Sayson, Beichman, Bryden, Ygouf, Gaspar, Thompson, Sanghi, Mawet, Greenbaum, Leisenring, Wolff, Rieke, \& Rieke}]{llopsayson2025searchingplanetsorbitingepsiloneridani}
Llop-Sayson, J., Beichman, C., Bryden, G., {et~al.} 2025, Searching for Planets Orbiting $\epsilon$~Eridani with JWST/NIRCam.
\newblock \doarXiv{2508.08463}

\bibitem[{{Lovis} {et~al.}(2017){Lovis}, {Snellen}, {Mouillet}, {Pepe}, {Wildi}, {Astudillo-Defru}, {Beuzit}, {Bonfils}, {Cheetham}, {Conod}, {Delfosse}, {Ehrenreich}, {Figueira}, {Forveille}, {Martins}, {Quanz}, {Santos}, {Schmid}, {S{\'e}gransan}, \& {Udry}}]{2017A&A...599A..16L}
{Lovis}, C., {Snellen}, I., {Mouillet}, D., {et~al.} 2017, \aap, 599, A16, \dodoi{10.1051/0004-6361/201629682}

\bibitem[{Lu {et~al.}(2024)Lu, Li, Cassese, \& Lin}]{lu2024dynamicalhistoryhip41378f}
Lu, T., Li, G., Cassese, B., \& Lin, D. 2024, The Dynamical History of HIP-41378 f -- Oblique Exorings Masquerading as a Puffy Planet.
\newblock \doarXiv{2410.00641}

\bibitem[{Marley {et~al.}(2021)Marley, Saumon, Morley, Fortney, Visscher, Freedman, \& Lupu}]{Marley2021}
Marley, M., Saumon, D., Morley, C., {et~al.} 2021, Sonora Bobcat: cloud-free, substellar atmosphere models, spectra, photometry, evolution, and chemistry, Sonora Bobcat,  Zenodo, \dodoi{10.5281/zenodo.5063476}

\bibitem[{{Mawet} {et~al.}(2019){Mawet}, {Hirsch}, {Lee}, {Ruffio}, {Bottom}, {Fulton}, {Absil}, {Beichman}, {Bowler}, {Bryan}, {Choquet}, {Ciardi}, {Christiaens}, {Defr{\`e}re}, {Gomez Gonzalez}, {Howard}, {Huby}, {Isaacson}, {Jensen-Clem}, {Kosiarek}, {Marcy}, {Meshkat}, {Petigura}, {Reggiani}, {Ruane}, {Serabyn}, {Sinukoff}, {Wang}, {Weiss}, \& {Ygouf}}]{2019AJ....157...33M}
{Mawet}, D., {Hirsch}, L., {Lee}, E.~J., {et~al.} 2019, \aj, 157, 33, \dodoi{10.3847/1538-3881/aaef8a}

\bibitem[{{Mennesson} {et~al.}(2022){Mennesson}, {Bailey}, {Zellem}, {Hildebrandt}, {Ygouf}, {Rhodes}, {Zimmerman}, {Nemati}, {Gonzalez}, {Cady}, {Kern}, {Koch}, {Krist}, {Heydorff}, {Luchik}, {Mok}, {Morrissey}, {Poberezhskiy}, {Riggs}, {Shi}, {Zhao}, {Akeson}, {Armus}, {Greenbaum}, {Ingalls}, \& {Lowrance}}]{2022SPIE12180E..1WM}
{Mennesson}, B., {Bailey}, V.~P., {Zellem}, R., {et~al.} 2022, in Society of Photo-Optical Instrumentation Engineers (SPIE) Conference Series, Vol. 12180, Space Telescopes and Instrumentation 2022: Optical, Infrared, and Millimeter Wave, ed. L.~E. {Coyle}, S.~{Matsuura}, \& M.~D. {Perrin}, 121801W, \dodoi{10.1117/12.2629176}

\bibitem[{Miller {et~al.}(2023)Miller, Filacchione, Cuzzi, Nicholson, Hedman, Baillie, Johnson, Tseng, Estrada, Waite, Ciarniello, Ferrari, Zhang, Hendrix, \& Moses}]{miller2023compositionsaturnsrings}
Miller, K.~E., Filacchione, G., Cuzzi, J., {et~al.} 2023, The composition of Saturn's rings.
\newblock \doarXiv{2311.17344}

\bibitem[{{Mosqueira} \& {Estrada}(2003)}]{2003Icar..163..198M}
{Mosqueira}, I., \& {Estrada}, P.~R. 2003, \icarus, 163, 198, \dodoi{10.1016/S0019-1035(03)00076-9}

\bibitem[{{National Academies of Sciences, Engineering, and Medicine}(2021)}]{2021pdaa.book.....N}
{National Academies of Sciences, Engineering, and Medicine}. 2021, {Pathways to Discovery in Astronomy and Astrophysics for the 2020s}, \dodoi{10.17226/26141}

\bibitem[{{Nayak} {et~al.}(2017){Nayak}, {Lupu}, {Marley}, {Fortney}, {Robinson}, \& {Lewis}}]{2017PASP..129c4401N}
{Nayak}, M., {Lupu}, R., {Marley}, M.~S., {et~al.} 2017, \pasp, 129, 034401, \dodoi{10.1088/1538-3873/129/973/034401}

\bibitem[{{Ohno} \& {Fortney}(2022)}]{2022ApJ...930...50O}
{Ohno}, K., \& {Fortney}, J.~J. 2022, \apj, 930, 50, \dodoi{10.3847/1538-4357/ac6029}

\bibitem[{Ohno {et~al.}(2022)Ohno, Thao, Mann, \& Fortney}]{Ohno_2022}
Ohno, K., Thao, P.~C., Mann, A.~W., \& Fortney, J.~J. 2022, The Astrophysical Journal Letters, 940, L30, \dodoi{10.3847/2041-8213/ac9f3f}

\bibitem[{Piro \& Vissapragada(2020)}]{Piro_2020}
Piro, A.~L., \& Vissapragada, S. 2020, The Astronomical Journal, 159, 131, \dodoi{10.3847/1538-3881/ab7192}

\bibitem[{Sanghi {et~al.}(2026)Sanghi, Mang, Llop-Sayson, Mamajek, Thompson, Sur, Beichman, Bryden, Gaspar, Leisenring, Mawet, Morley, Ruffio, Wolff, \& Ygouf}]{sanghi2026worldsdooriiiindirect}
Sanghi, A., Mang, J., Llop-Sayson, J., {et~al.} 2026, Worlds Next Door. III. Indirect Evidence for Enhanced Atmospheric Metallicity and/or the Presence of Water Clouds in the Nearest Jupiter-analog $\epsilon$ Eri b.
\newblock \doarXiv{2602.23423}

\bibitem[{Savransky \& Grigoryan(2025)}]{corgietc}
Savransky, D., \& Grigoryan, H. 2025, {corgietc}, 1.3.0,  Github.
\newblock \url{https://github.com/roman-corgi/corgietc}

\bibitem[{Smith {et~al.}(1979)Smith, Soderblom, Johnson, Ingersoll, Collins, Shoemaker, Hunt, Masursky, Carr, Davies, Cook, Boyce, Danielson, Owen, Sagan, Beebe, Veverka, Strom, McCauley, Morrison, Briggs, \& Suomi}]{doi:10.1126/science.204.4396.951}
Smith, B.~A., Soderblom, L.~A., Johnson, T.~V., {et~al.} 1979, Science, 204, 951, \dodoi{10.1126/science.204.4396.951}

\bibitem[{Smith {et~al.}(1989)Smith, Soderblom, Banfield, c.~Barnet, Basilevsky, Beebe, Bollinger, Boyce, Brahic, Briggs, Brown, c.~Chyba, s.~A.~Collins, Colvin, Cook, Crisp, Croft, Cruikshank, Cuzzi, Danielson, Davies, Jong, Dones, Godfrey, Goguen, Grenier, Haemmerle, Hammel, c.~J.~Hansen, c.~P.~Helfenstein, Howell, Hunt, Ingersoll, Johnson, Kargel, Kirk, Kuehn, Limaye, Masursky, McEwen, Morrison, Owen, Owen, Pollack, c.~c. Porco, Rages, Rogers, Rudy, Sagan, Schwartz, Shoemaker, Showalter, Sicardy, Simonelli, Spencer, Sromovsky, Stoker, Strom, Suomi, Synott, Terrile, Thomas, Thompson, Verbiscer, \& Veverka}]{doi:10.1126/science.246.4936.1422}
Smith, B.~A., Soderblom, L.~A., Banfield, D., {et~al.} 1989, Science, 246, 1422, \dodoi{10.1126/science.246.4936.1422}

\bibitem[{{Spergel} {et~al.}(2015){Spergel}, {Gehrels}, {Baltay}, {Bennett}, {Breckinridge}, {Donahue}, {Dressler}, {Gaudi}, {Greene}, {Guyon}, {Hirata}, {Kalirai}, {Kasdin}, {Macintosh}, {Moos}, {Perlmutter}, {Postman}, {Rauscher}, {Rhodes}, {Wang}, {Weinberg}, {Benford}, {Hudson}, {Jeong}, {Mellier}, {Traub}, {Yamada}, {Capak}, {Colbert}, {Masters}, {Penny}, {Savransky}, {Stern}, {Zimmerman}, {Barry}, {Bartusek}, {Carpenter}, {Cheng}, {Content}, {Dekens}, {Demers}, {Grady}, {Jackson}, {Kuan}, {Kruk}, {Melton}, {Nemati}, {Parvin}, {Poberezhskiy}, {Peddie}, {Ruffa}, {Wallace}, {Whipple}, {Wollack}, \& {Zhao}}]{2015arXiv150303757S}
{Spergel}, D., {Gehrels}, N., {Baltay}, C., {et~al.} 2015, arXiv e-prints, arXiv:1503.03757, \dodoi{10.48550/arXiv.1503.03757}

\bibitem[{Sucerquia {et~al.}(2020)Sucerquia, Alvarado-Montes, Zuluaga, Montesinos, \& Bayo}]{Sucerquia_2020}
Sucerquia, M., Alvarado-Montes, J.~A., Zuluaga, J.~I., Montesinos, M., \& Bayo, A. 2020, Monthly Notices of the Royal Astronomical Society: Letters, 496, L85–L90, \dodoi{10.1093/mnrasl/slaa080}

\bibitem[{Teachey(2024)}]{teachey2024detectingcharacterizingexomoonsexorings}
Teachey, A. 2024, Detecting and Characterizing Exomoons and Exorings (Handbook of Exoplanets, 2nd Edition).
\newblock \doarXiv{2401.13293}

\bibitem[{{Thompson} {et~al.}(2025){Thompson}, {Nielsen}, {Ruffio}, {Blunt}, \& {Marois}}]{2025AJ....170..301T}
{Thompson}, W., {Nielsen}, E.~L., {Ruffio}, J.-B., {Blunt}, S., \& {Marois}, C. 2025, \aj, 170, 301, \dodoi{10.3847/1538-3881/ae0cbd}

\bibitem[{{Tschudi} {et~al.}(2024){Tschudi}, {Schmid}, {Nowak}, {Le Coroller}, {Hunziker}, {van Holstein}, {Perrot}, {Mouillet}, {Augereau}, {Bazzon}, {Beuzit}, {Boccaletti}, {Bonse}, {Chauvin}, {Desidera}, {Dohlen}, {Dominik}, {Engler}, {Feldt}, {Girard}, {Gratton}, {Henning}, {Kasper}, {Kervella}, {Lagrange}, {Langlois}, {Martinez}, {M{\'e}nard}, {Meyer}, {Milli}, {Pragt}, {Puget}, {Quanz}, {Roelfsema}, {Sauvage}, {Szul{\'a}gyi}, {Thalmann}, \& {Zurlo}}]{2024A&A...687A..74T}
{Tschudi}, C., {Schmid}, H.~M., {Nowak}, M., {et~al.} 2024, \aap, 687, A74, \dodoi{10.1051/0004-6361/202449442}

\bibitem[{Veenstra {et~al.}(2025)Veenstra, Zuluaga, Alvarado-Montes, Sucerquia, \& Stam}]{Veenstra_2025}
Veenstra, A.~K., Zuluaga, J.~I., Alvarado-Montes, J.~A., Sucerquia, M., \& Stam, D.~M. 2025, Astronomy \& Astrophysics, 693, A310, \dodoi{10.1051/0004-6361/202347194}

\bibitem[{{Wang} \& {The Roman Coronagraph CPP Team}(2024)}]{2024ESS.....520203W}
{Wang}, J., \& {The Roman Coronagraph CPP Team}. 2024, in AAS/Division for Extreme Solar Systems Abstracts, Vol.~56, AASTCS10, Extreme Solar Systems V, 202.03

\bibitem[{{Winn} \& {Fabrycky}(2015)}]{2015ARA&A..53..409W}
{Winn}, J.~N., \& {Fabrycky}, D.~C. 2015, \araa, 53, 409, \dodoi{10.1146/annurev-astro-082214-122246}

\bibitem[{Wisdom {et~al.}(2022)Wisdom, Dbouk, Militzer, Hubbard, Nimmo, Downey, \& French}]{doi:10.1126/science.abn1234}
Wisdom, J., Dbouk, R., Militzer, B., {et~al.} 2022, Science, 377, 1285, \dodoi{10.1126/science.abn1234}

\bibitem[{{Wolff} {et~al.}(2024){Wolff}, {Wang}, {Stapelfeldt}, {Bailey}, {Savransky}, {Hom}, {Biller}, {Brandner}, {Anche}, {Blunt}, {Brinjikji}, {Girard}, {Krause}, {Li}, {Livingston}, {Millar-Blanchaer}, {Noel}, {Pueyo}, {De Rosa}, {Samland}, \& {Schragal}}]{2024SPIE13092E..55W}
{Wolff}, S.~G., {Wang}, J., {Stapelfeldt}, K., {et~al.} 2024, in Society of Photo-Optical Instrumentation Engineers (SPIE) Conference Series, Vol. 13092, Space Telescopes and Instrumentation 2024: Optical, Infrared, and Millimeter Wave, ed. L.~E. {Coyle}, S.~{Matsuura}, \& M.~D. {Perrin}, 1309255, \dodoi{10.1117/12.3020205}

\bibitem[{Zellem {et~al.}(2022)Zellem, Nemati, Gonzalez, Ygouf, Bailey, Cady, Colavita, Hildebrandt, Maier, Mennesson, Payne, Zimmerman, Belikov, Rosa, Debes, Douglas, Girard, Groff, Kasdin, Lowrance, Macintosh, Ryan, \& Weisberg}]{zellem2022nancygraceromanspace}
Zellem, R.~T., Nemati, B., Gonzalez, G., {et~al.} 2022, Nancy Grace Roman Space Telescope Coronagraph Instrument Observation Calibration Plan.
\newblock \doarXiv{2202.05923}

\bibitem[{{Zhu} \& {Dong}(2021)}]{2021ARA&A..59..291Z}
{Zhu}, W., \& {Dong}, S. 2021, \araa, 59, 291, \dodoi{10.1146/annurev-astro-112420-020055}

\bibitem[{Zuluaga {et~al.}(2022)Zuluaga, Sucerquia, \& Alvarado-Montes}]{ZULUAGA2022100623}
Zuluaga, J., Sucerquia, M., \& Alvarado-Montes, J. 2022, Astronomy and Computing, 40, 100623, \dodoi{https://doi.org/10.1016/j.ascom.2022.100623}

\bibitem[{{Zuluaga} {et~al.}(2015){Zuluaga}, {Kipping}, {Sucerquia}, \& {Alvarado}}]{2015ApJ...803L..14Z}
{Zuluaga}, J.~I., {Kipping}, D.~M., {Sucerquia}, M., \& {Alvarado}, J.~A. 2015, \apjl, 803, L14, \dodoi{10.1088/2041-8205/803/1/L14}

\end{thebibliography}
\bibliographystyle{aasjournal}
\end{document}